\documentclass[11pt,reqno]{amsart}

\usepackage{enumerate}
\usepackage{dsfont}
\usepackage{bbm}
\usepackage{color}
\usepackage{a4wide}
\usepackage[all]{xy}
\usepackage[colorlinks=true,linkcolor=blue]{hyperref}
\usepackage{parskip}
\usepackage[framemethod=tikz]{mdframed}

\hypersetup{%
    ,urlcolor=blue
    ,citecolor=red
    ,linkcolor=blue
    }

\allowdisplaybreaks

\definecolor{mycolor}{rgb}{0.122, 0.435, 0.698}

\newmdenv[innerlinewidth=0.5pt, roundcorner=4pt,linecolor=mycolor,innerleftmargin=6pt,
innerrightmargin=6pt,innertopmargin=6pt,innerbottommargin=6pt]{mybox}

\makeatletter
\renewcommand*{\eqref}[1]{%
  \hyperref[{#1}]{\textup{\tagform@{\ref*{#1}}}}%
}
\makeatother

\def\1{\mathbbm{1}}
\newtheorem{theorem}{Theorem}[section]

\theoremstyle{definition}

\theoremstyle{remark}

\numberwithin{equation}{section}
\usepackage{comment} % Block comment
\usepackage[utf8]{inputenc} % allow utf-8 input
\usepackage[T1]{fontenc}    % use 8-bit T1 .
\usepackage{hyperref}       % hyperlinks
\usepackage{url}            % simple URL typesetting
\usepackage{booktabs}       % professional-quality tables
\usepackage{amsfonts}       % blackboard math symbols
\usepackage{amsmath}        % maths
\usepackage{amsthm}         % maths
\usepackage{nicefrac}       % compact symbols for 1/2, etc.
\usepackage{microtype}      % microtypography
\usepackage{graphicx}
\usepackage{doi}
\usepackage{amsmath}
\usepackage{float}
\usepackage{enumitem} % Compact Itemize
\usepackage{multirow}
\usepackage{breqn} % Break line

\usepackage{natbib} % For alphabetical order of references
\usepackage{stmaryrd} % \llbracket

\usepackage{listofitems} % for \readlist to create arrays
\usepackage{tikz}
\usepackage{etoolbox} % for \ifthen
\usepackage{listofitems} % for \readlist to create arrays
\usetikzlibrary{arrows.meta} % for arrow size
\usepackage[outline]{contour} % glow around text
\contourlength{1.4pt}

\tikzset{>=latex} % for LaTeX arrow head
\usepackage{xcolor}
\colorlet{myred}{red!80!black}
\colorlet{myblue}{blue!80!black}
\colorlet{mygreen}{green!60!black}
\colorlet{myorange}{orange!70!red!60!black}
\colorlet{mydarkred}{red!30!black}
\colorlet{mydarkblue}{blue!40!black}
\colorlet{mydarkgreen}{green!30!black}
\tikzstyle{mynode}=[thick,draw=blue,fill=blue!20]
\tikzstyle{node}=[very thick,circle,draw=myblue,minimum size=22,inner sep=0.5,outer sep=0.6]
\tikzstyle{node in}=[node,green!15!black,draw=mygreen,fill=mygreen!25]
\tikzstyle{node hidden}=[node,blue!15!black,draw=myblue,fill=myblue!20]
\tikzstyle{node convol}=[node,orange!15!black,draw=myorange,fill=myorange!20]
\tikzstyle{node out}=[node,red!15!black,draw=myred,fill=myred!20]
\tikzstyle{connect}=[thick,mydarkblue] %,line cap=round
\tikzstyle{connect arrow}=[-{Latex[length=4,width=3.5]},thick,mydarkblue,shorten <=0.5,shorten >=1]
\tikzset{ % node styles, numbered for easy mapping with \nstyle
  node 1/.style={node in},
  node 2/.style={node hidden},
  node 3/.style={node out},
}
\def\nstyle{int(\lay<\Nnodlen?min(2,\lay):3)} % map layer number onto 1, 2, or 3
\hypersetup{
pdftitle={Deep Calibration of Interest Rate Models},
pdfsubject={Rates Models, Calibration, Deep Learning},
pdfauthor={Djibril SARR},
pdfkeywords={Interest Rate Models, G2++, CIR, Calibration, Neural Networks},
}

\begin{document}
	\newpage
	\title[Deep Calibration of Interest Rate Models]{Deep Calibration of Interest Rate Models}
	
	%    Remove any unused author tags.
	
	%    author 2 information
\author{Mohamed BEN ALAYA}
\address{Mohamed Ben Alaya, Laboratoire De Math\'ematiques Rapha{\"{e}}l Salem, UMR 6085, Universit\'e De Rouen, Avenue de L'Universit\'e Technop\^ole du Madrillet, 76801 Saint-Etienne-Du-Rouvray, France}
\curraddr{}
\email{mohamed.ben-alaya@univ-rouen.fr}
\thanks{}

%    author 2 information
\author{Ahmed KEBAIER}
\address{Ahmed Kebaier, Université d'Evry, LAMME, UMR 8071,  23 Bd. de France, 91037 Évry Cedex, France}
\curraddr{}
\email{ahmed.kebaier@univ-evry.fr}

%    author 1 information
\author{Djibril SARR}
\address{Djibril Sarr, Université Sorbonne Paris Nord, LAGA, CNRS, UMR 7539,  F-93430, Villetaneuse, France and FBH Associés, 11 Rue du 4 septembre, 75002 Paris}
\curraddr{}
\email{sarr@math.univ-paris13.fr, djibril.sarr@fbh-associes.com}

	\subjclass[2020]{91G30, 91G05, 91G70, 68T07, 60G15}
	
	\keywords{Interest Rate models, Calibration, G2++, Intensity Models, Neural Networks, Deep Learning}
	
	\date{11/10/2022}
	
%	\dedicatory{LAGA, Universit\'e Paris 13 }
	\maketitle

\begin{abstract}
    For any financial institution, it is essential to understand the behavior of interest rates. Despite the growing use of Deep Learning, for many reasons (expertise, ease of use, etc.), classic rate models such as CIR and the Gaussian family are still widely used. In this paper, we propose to calibrate the five parameters of the G2++ model using Neural Networks. Our first model is a Fully Connected Neural Network and is trained on covariances and correlations of Zero-Coupon and Forward rates. We show that covariances are more suited to the problem than correlations due to the effects of the unfeasible backpropagation phenomenon, which we analyze in this paper. The second model is a Convolutional Neural Network trained on Zero-Coupon rates with no further transformation. Our numerical tests show that our calibration based on deep learning outperforms the classic calibration method used as a benchmark. Additionally, our Deep Calibration approach is designed to be systematic. To illustrate this feature, we applied it to calibrate the popular CIR intensity model.
    
\end{abstract}

\section{Introduction}
\label{sec:SOA}
Governments, industries, and banks have to manage the behavior of various financial indicators. Whether it is to manage risks or optimize investment returns, it is necessary to understand, forecast, and stress drivers of fair value assets, for example. Among those drivers are interest rates (IR), which can affect IR derivatives such as IR swaps, swaptions, cross-currency swaps, and so on. IR models are widely used in banks and financial institutions to assess IR behavior. There are many models. The Vasicek model paved the way for these models \citep{vasicek1977equilibrium} and was later improved by many others. The Cox-Ingersoll-Ross Model (CIR) \citep{CIR_base} is often used as it is quite simple to use and calibrate. However, it is unifactorial, which limits its use. The Gaussian model G2++ allows for two factors as it is a deterministically shifted two-factor Vasicek model.

This document focuses on the calibration of the G2++ model (also called the Gauss2++ or Gaussian two-factor model) using deep learning (DL) techniques. More precisely, we will introduce two different approaches applied to different sets of relevant data to calibrate the G2++ model using Neural Networks (NN). These approaches are expected to be at least as accurate as the \emph{classic} ones (see the next paragraphs of the current section for the state of the art), to perform faster, and above all, to be easier to use regarding the treatments necessary on the data. In what follows, we will present two different approaches to calibrate IR models.

We consider a time horizon $T>0$. We consider a probability space $\left(\Omega, \mathcal{F}, (\mathcal{F}_t)_{t \in [0,T]}, \mathbb{Q}\right)$. The sample space $\Omega$ represents the set of all possible outcomes. $\mathcal{F}$ is a $\sigma$-field representing the set of events $A \subset \Omega$. $\mathbb{Q}$ is the risk-neutral probability that assigns a probability to each event in $\mathcal{F}$ and under which discounted obligation prices are martingales. In this setup, we recall the equations defining the G2++ model:
\begin{equation}
    \; \\
    \begin{aligned}
        & dr(t) = \phi(t) + x(t) + y(t) \\
        & dx(t) = -K_xx(t)dt + \sigma_xdW_t^x, \; x(0)=x_0 \\
        & dy(t) = -K_yy(t)dt + \sigma_ydW_t^y, \; y(0)=y_0 \; \text{with} \; K_x, K_y, \sigma_x, \sigma_y > 0.
    \end{aligned} 
    \\
    \label{eq:G2pp}
\end{equation}
In the above equation, $(W_t^x, W_t^y)_{t \in [0,T]}$ is a two-dimensional $(\mathcal{F}_t)_{t \in [0,T]}$-Brownian motion with correlation $\rho$, i.e., $[dW_t^x, dW_t^y] = \rho \, dt$. The function $\phi(t)$ is, as usual, deterministic and allows the model to fit the initial term structure of forward rates. In the remainder of this work, the function 
$
\phi(T) = f^{Market}(0,T) + \frac{\sigma_x^2}{2K_x^2} \left(1 - e^{-K_x T}\right)^2 + \frac{\sigma_y^2}{2K_y^2} \left(1 - e^{-K_y T}\right)^2 + \rho\frac{ \sigma_x \sigma_y}{K_x K_y} \left(1 - e^{-K_x T}\right) \left(1 - e^{-K_y T}\right),
$
where $f^{Market}(0,T)$ is the market initial forward rate (see Section 4.2.1 of \cite{brigo2006interest} for more details). We then require the $(x, y)$-model to fit the market \textit{as best as possible} via calibration and then update the function $\phi(t)$ using the optimal parameters (see discussion after Lemma 2 of \cite{mbaye2022affine}). The scalars $K_x, K_y, \sigma_x, \sigma_y$ are positive parameters, and $\rho$ is in $[-1,1]$. These are the parameters of the model and require calibration.

Despite our work being applied to G2++, the approaches are actually systematic, meaning that they are not model-dependent as long as some requirements, which will be detailed in Section \ref{sec:Model}, are met. In Section \ref{sec:apdx_CIR}, we show another possible application with the CIR intensity model. Hence, the Deep Calibration process we introduce here is a contribution to the calibration of IR models and to the application of DL to finance. The first matter is a historical one, and many solutions have been introduced over the years.

Indeed, the accuracy of the model used for IR is highly dependent on the quality of the calibration. For the one-factor Hull-White (HW) model, \cite{gurrieri2009calibration} defines the calibration by these three elements:
\begin{enumerate}
    \item The choice of constant or time-dependent mean reversion and volatility.
    \item The choice of products to calibrate to, and whether to calibrate locally or globally.
    \item Whether to optimize the mean reversion and the volatility together or separately, and in the latter case, how to estimate one independently of the other.
\end{enumerate}
We can generalize this definition by extending the first and third points to include not only the mean reversion and volatility but any parameters required in the model. Each of these three points is important and will be discussed in a moderate to thorough manner.

Independently of these points, in general, the calibration process involves minimizing the distance between a theoretical value obtained from the model and the actual corresponding market observation. \cite{hull2001general}, for example, uses numerical optimization techniques such as the Levenberg-Marquardt algorithm to find the set of volatility parameters that minimize the error between the model's prices for cap and floor options and swaptions. The error used is the sum of the squares of the differences (SSD), that is,
$
    \theta^{*} = \underset{\theta}{\mathrm{argmin}} \sum_{i=1}^N{\left(OP_i(\theta) - OP_i^{market}\right)^2}
$
where $OP(\cdot)$ defines an option price, $N$ is the number of options included in the \emph{calibration basket}, and $\theta$ is the set of parameters of the model to calibrate. The same idea of an optimization process on swaption prices can also be seen in \cite{russo2019calibration}, where, instead of using the SSD, the relative error is used, that is,
$
    \theta^{*} = \underset{\theta}{\mathrm{argmin}} \sqrt{\sum_{i=1}^N{\left(\frac{OP_i(\theta) - OP_i^{market}}{OP_i^{market}}\right)^2}}.
$
\cite{schlenkrich2012efficient} also calibrates the HW model on swaptions. The optimization process is done using Gauss-Newton and Adjoint Broyden quasi-Newton methods. The required derivatives are approximated by automatic differentiation techniques.

Other calibration methods address more specific and advanced questions. For instance, in a macroeconomic context with low to negative IRs, the methods above may not be well suited. \cite{russo2017calibrating} addresses the problem of negative IRs with a focus on the second point of our definition, the choice of products to calibrate to. More precisely, while they still use swaptions, they aim at finding the best swaption quotation. They conclude that calibrating the Hull-White model, the shift-extended CIR model, or the shift-extended squared Gaussian model using the shifted log-normal or normal volatility results in more stable quotations, which in turn provides more stable parameters.

\cite{orlando2019interest} tackle the calibration of the CIR model with negative or near-zero IR values. Their approach consists of translating the IRs to positive values, thereby retaining the initial volatility. The rate is shifted by a constant $\alpha$ (allowing one to keep the initial dynamics), i.e.,
$
    r_{shift}(t) = r_{real}(t) + \alpha.
$
They use the 99th percentile of the empirical IR probability distribution as the constant. For the calibration process, instead of minimizing the error between theoretical and market data, the authors take advantage of the fact that parameters' likelihoods in the CIR model are known (for example, \cite{kladivko2007maximum, ben2013asymptotic}).

Regarding the second matter, the use of DL in finance, DL techniques, especially neural networks (NNs), are increasingly being applied in various fields, including finance, where many applications have been found. According to \cite{huang2020deep}, for example, feedforward neural networks (FNNs), which consist of multiple layers of fully connected perceptrons \citep{pal1992multilayer}, and Long Short-Term Memory (LSTM) networks \citep{hochreiter1997long}, which are recurrent networks (NNs that can process information in two directions instead of one like FNNs), are the architectures mainly used in forecasting (exchange rates, option prices, risks, stock markets, etc.). 

Additionally, DL can be used to replicate financial instruments or characteristic behaviors. \cite{heaton2016deep}, for example, use autoencoders \cite{ng2011sparse}, which are neural architectures originally designed to automatically learn features from unlabeled data to replicate the inputs. The authors' purpose is to replicate indexes such as the S\&P 500. A similar idea of replication is found in \cite{bloch2019neural}, where the authors use a neural network with dimensionality reduction through Principal Component Analysis (PCA) to learn the dynamics of the implied volatility surface. Replicating the dynamics of indexes or volatilities can be a first step in choosing an investment or hedging strategy. However, DL can also be used directly for both of these purposes. Regarding investment strategies, \cite{heaton2017deep} introduces an approach called \emph{Deep Portfolios}, which aims to understand the key factors driving asset prices before generating the mean-variance efficient portfolio. As for hedging, \cite{buehler2019deep}, with the \emph{Deep Hedging} approach, uses reinforcement learning—which employs a reward system so the algorithm can learn from its experience (in-depth details can be found in \cite{kaelbling1996reinforcement})—to find the optimal hedge given available instruments in a hypothetical market driven by the Heston model.

Another application of DL in finance that is gaining attention is asset pricing. For example, \cite{chen2020deep}'s approach combines three different architectures with distinct purposes: an FNN aimed at understanding non-linearities, an LSTM network to identify sets of economic state processes that help the model understand macroeconomic conditions relevant to asset pricing, and a generative adversarial network (GAN) (for more on GANs, see \cite{creswell2018generative}) to identify the strategies with the most unexplained pricing information. The added value of the authors' work is also in their inclusion of the no-arbitrage assumption via a stochastic discount factor, which they demonstrate helps improve performance. \cite{jang2021deepoption} use a deep FNN to predict market option values. Their performance is enhanced by pre-training the network with data generated using parametric option pricing methods (Black-Scholes, Monte Carlo, finite differences, and binomial trees) to overcome the lack of market data and unbalanced datasets.

Finally, another application of DL to finance, which is the main focus of this article, is the calibration of financial models. \cite{pironneau2019calibration}, for example, calibrated the Heston model for European options using FNNs. While the author achieves good precision with a relatively shallow architecture, performance stalls at a certain point. \cite{hernandez2016model} also uses FNNs to calibrate financial models in general and applies this approach to the one-factor Hull-White model using swaption prices. The author outperforms classic methods in terms of calibration times. In \cite{horvath2021deep}, the authors calibrate the volatility surface using neural networks. They draw inspiration from implied volatility and option prices, conceptualizing them as a set of pixels. After achieving good performance in terms of computation time and accuracy, the authors analyze the use of deep calibration algorithms to identify the stochastic model that most accurately represents a given dataset, aiming to discern which model best characterizes the market.

Very recently, \cite{buchel2022deep} conducted a comparison of the calibration of an interest rate term structure model based on the Trolle and Schwartz volatility model, with and without the integration of artificial neural networks (NNs). Their findings indicate that employing NNs leads to faster and more stable results over time. Notably, their methodology utilizes NNs to learn the model itself. Consequently, to obtain the calibrated parameters, a final optimization step is required. The calibration framework they used is based on the work of \citep{liu2019neural}, where the authors use NNs to calibrate the Heston pricing model and the Bates model. In that paper, the authors also have an initial step that involves learning to map parameters to prices (referred to as the forward pass in both references) before proceeding to the actual calibration. In comparison, our approach involves a neural network architecture that directly outputs the parameters of the models, thereby completing the calibration process directly.

The approach we present differs from the aforementioned calibration processes using DL in both the input data (we do not use options prices) and the architectures we propose. Indeed, depending on the input data, we use either a Feedforward Neural Network (FNN) or a Convolutional Neural Network (CNN) (see \citep{albawi2017understanding} for details). Notably, in the previously mentioned works, FNNs are primarily used. Overall, one of the primary distinctions between the research presented in this paper and much of the existing literature on the subject, is that we focus directly on interest rate models and their parameters instead of pricing functions. Additionally, this work introduces two types of calibration: direct and indirect calibration.

Moreover, in this paper, as part of our efforts to assess the robustness of the calibration process, we conduct an analysis where we deliberately introduce noise into the inputs. Additionally, in our systematic approach, the models we introduce are applied to the simultaneous calibration of the five parameters of the G2++ model but can be applied to any model, as long as some non-restrictive conditions, detailed in Section \ref{sec:Model}, are met. Our work also differs from the current literature regarding the types of inputs since we train our first neural network on covariance and correlation matrices of Zero-Coupon (ZC) rates and forward (FWD) rates. We show that for feedforward calibration neural networks, covariances provide richer information compared to correlations. To our knowledge, feeding neural networks using covariance matrices to calibrate models has not yet been explored in the literature, and our paper fills this gap. This leads us to the introduction of the unfeasible backpropagation phenomenon (see Theorem \ref{theo:unfeasbackprop}), which supports the choice of covariances over correlations—a concept that, to the best of our knowledge, had not been previously addressed in existing literature.

In what follows, Section \ref{sec:Model} will describe the models we introduce in this article. For each model, we will also explain how the dataset used for training was constructed. The results obtained, both in terms of accuracy and computational performance, will be presented in Section \ref{sec:resec}. Interesting intermediary results are also provided in this section, where we demonstrate that covariances of ZC rates are more suited for the Deep Calibration process than correlations due to vanishing gradients. More specifically, we present a theorem explaining when backpropagation becomes unfeasible (see Theorem \ref{theo:unfeasbackprop}), a result that, to our knowledge, is not commonly found in the literature. Section \ref{sec:compsec} will compare our results to classic calibration methods, showing better accuracy and a significant reduction in calibration time. Finally, Section \ref{sec:apdx_CIR} will illustrate how the work on the G2++ model can be easily applied to a CIR intensity model.

\section{A systematic approach with Deep Calibration (DC) for interest rate models}
\label{sec:Model}

In what follows, we will present two different approaches to calibrate IR models. Both approaches are not model-dependent (and thus systematic) and could be applied to any model (financial or not) as long as what we call the Observable Quantity of Interest (OQOI), which must be a relevant information bearer, can be both observed (in the market, in our financial context) and obtained through an analytical expression from the model that one wants to calibrate. 

One of the main concerns of our DC models is the ease of use regarding data. We aim to use easily available market data that require minimal transformations. For the training of our DL models, we decided to use synthetic data based on numerical simulations in both approaches. In more detail, we generate the synthetic datasets from a set of parameters calibrated from real market data (see Sections \ref{sec:datasetconstr} and \ref{sec:dataZC}). There are many reasons in favor of this approach:

\begin{itemize}
    \item The dataset becomes of unlimited size.
    \item We can modify the data to be more representative of a specific context (a stressed one, for example), or we can decide to include many different contexts.
    \item Unlike the other calibration papers mentioned previously, we can measure our error directly vis-à-vis the real parameters.
    \item Our DL architectures gets to actually learn what \textit{drives} the IR model. This means that when performing out-of-sample calibration (outside of the learning process), the DL model will find the parameters that truly suit the given IR model. If, instead, real-world data were used, we would have found the parameters corresponding to the real-world data closer to the IR model. In other words, using synthetic data allows us to replicate the behavior of the model we are currently calibrating.
\end{itemize}

The last point is important as it also helps us stay consistent with our aim of calibrating IR models rather than forecasting IR curves, which is a different task that has also received attention in the literature (see, for example, \cite{oh2000using, jacovides2008forecasting, vela2013forecasting}). The main reasons to support this choice are that, as of today, professionals and researchers using IR forecasts have more mastery over IR models than they do over DL models, since the former have been widely used for many decades. Additionally, IR models have many advantages; for instance, it is quite straightforward to stress IR forecasts via classic models (through real-world simulations or alterations of parameters).

The following two subsections present each approach by explaining how the analytical expressions allowing synthetic data generation are obtained, how the dataset is constructed, and which architecture is chosen. Both approaches—in different ways—use the risk-neutral values of ZC and FWD rates.

We recall the expression of the ZC bond price $P(\cdot,\cdot)$ at time $t \in [0, T]$ and maturity $T$ under the G2++ model (described by equation \eqref{eq:G2pp}):
\begin{equation}
    \begin{aligned}
        P(t, T)=& \frac{P^{M}(0, T)}{P^{M}(0, t)} \exp \{\mathcal{A}(t, T)\}, \\
        \text{ with } \mathcal{A}(t, T)=& \frac{1}{2}[V(t, T)-V(0, T)+V(0, t)] - \frac{1-e^{-K_x(T-t)}}{K_x} x(t)-\frac{1-e^{-K_y(T-t)}}{K_y} y(t),
    \end{aligned}
    \label{eq:A}
\end{equation}
and $P^{M}(0, T)$, the initial ZC bond price at maturity $T$, and  $V(\cdot, \cdot)$ verifying:
\begin{equation}
    \begin{aligned}
    V(t, T)=& \frac{\sigma_x^{2}}{K_x^{2}}\left[T-t+\frac{2}{K_x} e^{-K_x(T-t)}-\frac{1}{2K_x} e^{-2K_x(T-t)}-\frac{3}{2 K_x}\right]\\
    &+\frac{\sigma_y^{2}}{K_y^{2}}\left[T-t+\frac{2}{K_y} e^{-K_y(T-t)}-\frac{1}{2K_y} e^{-2K_y(T-t)}-\frac{3}{2K_y}\right] \\
    &+2 \rho \frac{\sigma_x \sigma_y}{K_xK_y}\left[T-t+\frac{e^{-K_x(T-t)}-1}{K_x}+\frac{e^{-K_y(T-t)}-1}{K_y}-\frac{e^{-(K_x+K_y)(T-t)}-1}{K_x+K_y}\right].
    \end{aligned}
    \label{eq:VV}
\end{equation}

Since the ZC rate $Z(.,.)$ satisfies $P(t, T)=e^{-(T-t)Z(t, T)}$, we have:
\begin{equation}
    Z(t, T) = -\frac{1}{T-t}ln \left\{\frac{P^{M}(0, T)}{P^{M}(0, t)} \right\} -\frac{1}{T-t} \{\mathcal{A}(t, T)\}.
    \label{eq:Z}
\end{equation}

Finally, as for the FWD rate $f(., .)$, we have $f(t, T) =  -\frac{\partial}{\partial T}lnP(t, T)$, then:
\begin{equation}
    \label{eq:fwd}
    \begin{aligned}
        f(t, T) &= -\frac{\partial}{\partial T} \ln P(t, T)=-\frac{\partial}{\partial T} A(t, T)\\
        &=-\frac {\sigma_{x}^{2}}{2K_{x}^{2}}\left[1-2{e}^{-K_x(T-t)}+e^{-2 K_x(T-t)}\right]\\
        &+\frac{\sigma_{y}^{2}}{2K_{y}^{2}}\left[1-2 e^{-K_y(T-t)}+e^{-2 K_y(T-t)}\right]\\
        &+ \rho \frac{\sigma_{x}\sigma_{y}}{k_{x} k_{y}}\left[1-e^{-K_x(T-t)}+e^{-K_y(T-t)}-e^{-(K_x + K_y)(T-t)}\right]\\
        &+e^{-K_x(T-t)} x(t)+e^{-K_{y}(T-t)} y(t).
        \end{aligned}
\end{equation}

The proofs of these expressions are given in the fourth chapter of \citep{brigo2006interest}.

Let us also define, for the remainder of this document, the set of parameters to calibrate, $\theta = \{ \theta^i \}_{i \in \llbracket 1, 5 \rrbracket} = \{ K_x, \; K_y, \; \sigma_x, \; \sigma_y, \; \rho \}$.

\vspace{0.5cm}
\subsection{Indirect Deep Calibration of the G2++ model}
\subsubsection{Covariances and Correlations as OQOI}\mbox{}\\ 
For our first approach, we start with the observation that ZC and FWD rates are indeed easily observable in the market and that an analytical expression can be derived from most, if not all, IR models. However, if we want to maintain a simple Fully Connected Network (FCN) architecture, directly using them as our OQOI might be time-consuming and less efficient. We decide to use intermediary quantities (hence the term \emph{Indirect Deep Calibration}) that bear enough compressed information from the ZC and FWD rates curves to allow for efficient calibration. Good candidates for the OQOI could be the correlation and covariance matrices of ZC and FWD rates variations.

From equation \eqref{eq:fwd} we can re-write: 
\begin{equation}
    \label{eq:forward2}
        f(t, T) = \psi(t, T) \; +e^{-K_x(T-t)} x(t) \; +e^{-K_{y}(T-t)} y(t), \; t \in [0, T]
\end{equation}

with
\begin{equation*}
    \begin{cases}
        & x(t) = e^{-K_xt}x_0 + \sigma_xe^{-K_xt}\int_0^te^{K_xs}dW^x_s\\
        & y(t) = e^{-K_yt}y_0 + \sigma_ye^{-K_yt}\int_0^te^{K_ys}dW^y_s.\\
    \end{cases} 
\end{equation*}

As the integrands in the above equations are deterministic, the covariance and the Itô's bracket match and for two given tenors $T_i$ and $T_j \in [0, T]$, we have 
\begin{equation}
    \begin{aligned}
        \label{eq:dcovfwd}
            d{\rm Cov}(f(t, T_i), f(t, T_j)) &= \Big[\sigma_x^2 e^{-K_x(T_i+T_j - 2t)} + \sigma_y^2 e^{-K_y(T_i+T_j-2t)} +\\
         & \rho\sigma_x\sigma_y\left(e^{-K_xT_i-K_yT_j+(K_x+K_y)t} + e^{-K_yT_i-K_xT_j+(K_x+K_y)t}\right)\Big]dt.\\
     \end{aligned} 
\end{equation}

From now on, following \citep{brigo2006interest} (see chapter four) and adopting their notations, we rather focus on the so-called instantaneous covariances and correlations defined by the following equation:
\begin{align}
  \label{eq:covcor1}
    {\rm Cov}\left(df(t, T_i), d f(t, T_j)\right) &:= \notag\\ &\hspace{-3.5cm} 
\left[X(t, T_i)X(t, T_j) + Y(t, T_i)Y(t, T_j) + \rho \left(X(t, T_i)Y(t, T_j) + X(t, T_j)Y(t, T_i) \right) \right]dt \notag\\ 
    {\rm Cor}\left(d f(t, T_i), d f(t, T_j)\right)  &:= \notag\\ &\hspace{-3.5cm}\frac{\left[X(t, T_i)X(t, T_j) + Y(t, T_i)Y(t, T_j) + \rho \left(X(t, T_i)Y(t, T_j) + X(t, T_j)Y(t, T_i) \right) \right]}{\sqrt{\left[X^2(t, T_i) + Y^2(t, T_i) + 2\rho X(t, T_i)Y(t, T_i)\right].\left[X^2(t, T_j) + Y^2(t, T_j) + 2\rho X(t, T_j)Y(t, T_j)\right]}},
\end{align}
where $X(t,T) =\sigma_xe^{-K_x(T-t)}$ and $Y(t,T) =\sigma_ye^{-K_y(T-t)}$. Note that in practice, $df(t,T)$ is interpreted as $f(t+\varepsilon,T)-f(t,T)$ with $dt=\varepsilon$. Equation \eqref{eq:dcovfwd} defines covariances for FWD rates. We want a similar expression for ZC rates. One just needs to re-write \eqref{eq:Z} in a way that splits the stochastic and the deterministic parts of the equation
\begin{equation}
    Z(t, T) = \widetilde{\psi}(t, T)- \frac{1-e^{-K_x(T-t)}}{TK_x} x(t)-\frac{1-e^{-K_y(T-t)}}{TK_y} y(t),
    \label{eq:Z2}
\end{equation}
with 
\begin{equation*}
    \widetilde{\psi}(t, T) = -\frac{1}{T}\ln \left\{\frac{P^{M}(0, T)}{P^{M}(0, t)} \right\} -\frac{1}{2T}[V(t, T)-V(0, T)+V(0, t)].
\end{equation*}

In the same way as for the FWDs, we use \eqref{eq:Z2} to introduce the instantaneous covariances and correlations for ZCs defined by
\begin{comment}
\begin{equation}
    \begin{aligned}
        \label{eq:dcovz}
         {\rm Cov}(dZ(t, T_i), dZ(t, T_j)) &= \sigma_x^2 \frac{1 - e^{-K_x(T_i - t)}}{K_xT_i}\frac{1 - e^{-K_x(T_j - t)}}{K_xT_j}dt+\\
         & \sigma_y^2 \frac{1 - e^{-K_y(T_i - t)}}{K_yT_i}\frac{1 - e^{-K_y(T_j - t)}}{K_yT_j}dt +\\
         & \rho\sigma_x\sigma_y\left[\frac{1 - e^{-K_x(T_i - t)}}{K_xT_i}\frac{1 - e^{-K_y(T_j - t)}}{K_yT_j}+\frac{1 - e^{-K_x(T_j - t)}}{K_xT_j}\frac{1 - e^{-K_y(T_i - t)}}{K_yT_i}\right].dt\\
     \end{aligned} 
\end{equation}
\end{comment}
\begin{align}
  \label{eq:covcor2}
    {\rm Cov}\left(dZ(t, T_i), d Z(t, T_j)\right) &:= \notag\\ &\hspace{-3.5cm} \left[\widetilde{X}(t, T_i)\widetilde{X}(t, T_j) + \widetilde{Y}(t, T_i)\widetilde{Y}(t, T_j) + \rho \left(\widetilde{X}(t, T_i)\widetilde{Y}(t, T_j) + \widetilde{X}(t, T_j)\widetilde{Y}(t, T_i) \right) \right]dt \notag\\ 
    {\rm Cor}\left(d Z(t, T_i), d Z(t, T_j)\right)  &:= \notag\\ &\hspace{-3.5cm}\frac{\left[\widetilde{X}(t, T_i)\widetilde{X}(t, T_j) + \widetilde{Y}(t, T_i)\widetilde{Y}(t, T_j) + \rho \left(\widetilde{X}(t, T_i)\widetilde{Y}(t, T_j) + \widetilde{X}(t, T_j)\widetilde{Y}(t, T_i) \right) \right]}{\sqrt{\left[\widetilde{X}^2(t, T_i) + \widetilde{Y}^2(t, T_i) + 2\rho \widetilde{X}(t, T_i)\widetilde{Y}(t, T_i)\right].\left[\widetilde{X}^2(t, T_j) + \widetilde{Y}^2(t, T_j) + 2\rho \widetilde{X}(t, T_j)\widetilde{Y}(t, T_j)\right]}},
\end{align}
where $\widetilde{X}(t,T) =\sigma_x \frac{1-e^{-K_x(T-t)}}{K_xT}$ and $\widetilde{Y}(t,T) =\sigma_y \frac{1-e^{-K_y(T-t)}}{K_yT}$.

In what follows, COV-ZC, COV-FWD, COR-ZC, and COR-FWD will designate, in that order, covariances of ZC rates, covariances of FWD rates, correlations of ZC rates, and correlations of FWD rates. A fortiori, when a pair of maturities $(T_1, T_2)$ is appended, it will designate the function applied to that specific pair of maturities; e.g., COV-ZC(5Y, 7Y) is the covariance of the ZC rate of maturity 5Y and the ZC rate of maturity 7Y.

\subsubsection{Data set construction}\mbox{}\\ 
\label{sec:datasetconstr}
The dataset construction is done in five steps:
\begin{enumerate}
    \item Calibrating reference parameters on real-world market data for the G2++ model using MSE (Mean Squared Error) minimization. The MSE is given by:
    \begin{equation}
        \label{eq:MSE}
        \text{MSE} = \frac{1}{n} \sum_{i=1}^{n} \left( y_i - \hat{y}_i \right)^2,
    \end{equation}
    where, $ y_i $ is the actual value for the $i$th observation, $ \hat{y}_i $ is the predicted value for the $i$th observation and $ n $ is the total number of observations.

    \item Extending the reference parameters to define a range of acceptance and drawing the parameters within the corresponding intervals.
    \item Using equations \eqref{eq:covcor1} and \eqref{eq:covcor2} to compute the set of covariances and correlations for ZC and FWD rates.
    \item Choosing the best scaling transformations.
    \item Finally, splitting the dataset into training and test sets.
\end{enumerate}

\vspace{0.5cm}
\paragraph{\textbf{Step 1: Choice of reference parameters}}\mbox{} \\
Table \ref{table:refparams} shows the parameters obtained from a calibration on real market Euro ZC rates data ranging from June 2019 to November 2020. The calibration leading to these parameters is performed on ZC correlations and covariances as described in Section \ref{sec:compmethodo}. These reference parameters are obtained without any use of NNs and are only used to infer a wide range of realistic parameters that will be used to train our NNs.
\begin{table}[H]
    \centering % used for centering table
    \begin{tabular}{c c c c c} % centered columns (5 columns)
    \hline \hline  % inserts double horizontal line
    $K_x$ & $K_y$ & $\sigma_x$ & $\sigma_y$  & $\rho$ \\ [1.ex] % inserts table
    \hline \hline %inserts double horizontal lines
     & & & & \\
    0.07173132 & 0.08930784 & 0.09465584 & 0.094675523 & -0.999318 \\ [1.ex] % inserting body of the table
    \hline %inserts single line
    \end{tabular}
    \vspace{1.5mm}
    \caption{Reference parameters} % title of Table
    \label{table:refparams} % is used to refer this table in the text
\end{table}

\paragraph{\textbf{Step 2: Extension of reference parameters and random draws}}\mbox{} \\ 
The idea here is, for a given $\gamma > 0$, to define intervals for our parameters ranging from $\theta^i - \gamma \theta^i$ to $\theta^i + \gamma \theta^i$ for each reference parameter $\theta^i$, except for the correlation, where we choose to take $\rho$ in the interval $\left[-0.999318, \; 0.999318\right]$. In what follows, we take $\gamma = \frac{2}{3}$. We obtain the following intervals:

\begin{table}[H]
    \centering % used for centering table
    \begin{tabular}{c | c c c c c} % centered columns (5 columns)
    \hline \hline  % inserts double horizontal line
     & $K_x$ & $K_y$ & $\sigma_x$ & $\sigma_y$  & $\rho$ \\ [1.ex] % inserts table
    \hline \hline %inserts double horizontal lines
     & & & & \\
    MIN	& 0.02391044	& 0.02976928	& 0.03155195	& 0.03155851	& -0.999318\\ [1.ex] % inserting body of the table
    MAX	& 0.1195522	& 0.1488464	& 0.15775973	& 0.15779254	& 0.999318\\ [1.ex] % inserting body of the table
    \hline %inserts single line
    \end{tabular}
    \vspace{1.5mm}
    \caption{Range of parameters} % title of Table
    \label{table:rangeparams} % is used to refer this table in the text
\end{table}
We decide to draw $N = 10,000$ parameters uniformly from each interval.

\vspace{0.5cm}
\paragraph{\textbf{Step 3: Computing covariances and correlations for ZCs and FWDs}} \mbox{} \\
The only question here is which maturities are chosen for the computations. Since in the real market, especially for short terms, liquidity can affect the rate curves (see, e.g., \citep{covitz2007liquidity}), we decide to use only tenors from 1 year to 12 years with a 1-year step. The covariance and correlation matrices are our features. They are transformed into vectors by stacking each column of the lower triangular matrix without the diagonal for correlations. Hence, we obtain an $\mathbb{R}^{nf}$ array with $nf = 66$ for correlations and $nf = 78$ for covariances. For three sets of randomly selected parameters, Figure \ref{fig:cov} below shows the covariance of ZC rates, and Figure \ref{fig:cor} shows the correlations of FWD rates. Appendix \ref{sec:apdx_mrkt_covcor} shows similar correlations and covariances from market ZC ratess.
\begin{figure}[H]
    \centering
    \includegraphics[height=5.5cm]{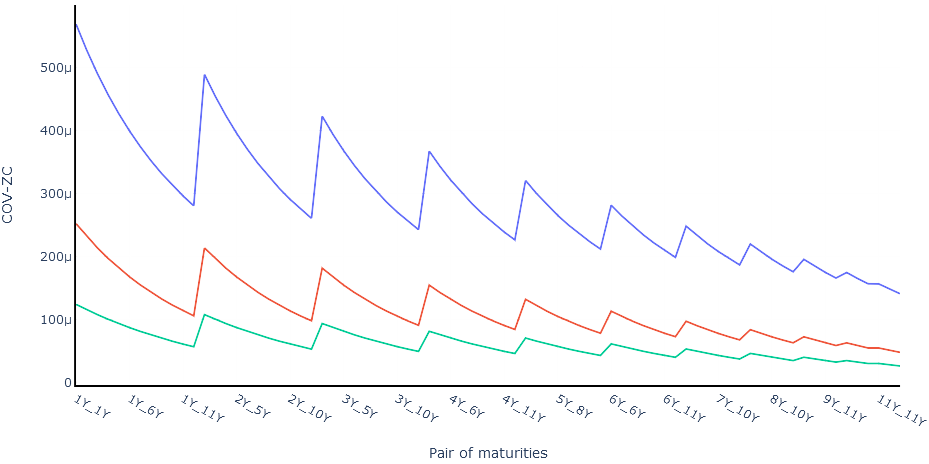} % second figure itself
    \caption{Covariances of ZC rates for random sets of parameters.}
    \label{fig:cov}
\end{figure}
\begin{figure}[H]
    \centering
    \includegraphics[height=5.5cm]{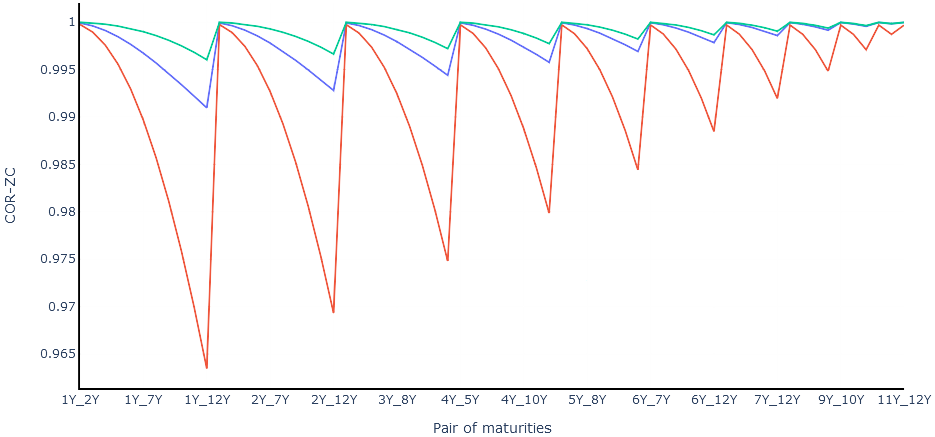} % first figure itself
    \caption{Correlations of FWD rates for random sets of parameters}
    \label{fig:cor}
\end{figure}
In the two graphics above, for the blue curve, $K_x=0.03$, $K_y=0.14$, $\sigma_x=0.1$, $\sigma_y=0.09$ and $\rho=0.84$; for the red curve, $K_x=0.04$, $K_y=0.1$, $\sigma_x=0.06$, $\sigma_y=0.1$ and $\rho=0.18$; finally for the green curve, $K_x=0.05$, $K_y=0.06$, $\sigma_x=0.05$, $\sigma_y=0.13$ and $\rho=-0.87$.

\vspace{0.5cm}
\paragraph{\textbf{Step 4: Scaling transformations}} \mbox{}\\
According to our numerical tests, the best transformation is obtained using the min-max scaler applied to both features and targets: $u_{\text{transformed}} = \frac{u - \min(u)}{\max(u) - \min(u)}.$

\vspace{0.5cm}
\paragraph{\textbf{Step 5: Splitting the data set}} \mbox{} \\ 
We now have at our disposal four different datasets (correlations and covariances for each ZC and FWD rates), each with 10,000 entries. For each of these four sets, we randomly allocate 80\% to the training set and 20\% to the test set. The training sets will be used for training the neural networks, while the test sets will be reserved for post-training analysis of our models' performance. Additionally, the test sets will be used to conduct comparisons between the results obtained from the different datasets.

\vspace{0.5cm}
\subsubsection{The neural network architecture}\mbox{}\\
As mentioned previously, we aim to keep a simple FCN architecture. Our neural network has five linear layers: one input layer with $nf$ neurons (which corresponds to the number of features—66 for correlations or 78 for covariances, as explained above), three hidden layers, the first with $h_1 = 1\,000$ neurons, the second with $h_2 = 1\,500$, and the last hidden layer with $h_3 = 1\,000$ neurons, and finally, one output layer with $np$ neurons (the number of parameters to calibrate, where $np = 5$). The first three layers are activated using a ReLU function, and the last one, as a prediction layer, has no activation. To help prevent overfitting, we also include a dropout with a probability of 0.25 between the last hidden layer and the prediction layer. 

Note that while artificial neural networks are susceptible to overfitting issues, these concerns may be mitigated, especially when employing synthetic data that allows for the creation of vast training sets without constraints, and when the data is devoid of sample noise due to simulation via deterministic functions of the parameters to calibrate (see \cite{buchel2022deep}). However, in our study, despite the ability to simulate an extensive range of inputs, we have opted to limit the dataset to 10\,000 entries, with only 80\% utilized for training purposes. This decision has been made to accelerate training procedures and reduce the number of epochs required. Furthermore, it is worth noting that in Section \ref{sec:compsec}, we intentionally introduce noise to the inputs to evaluate the resilience of our model. Therefore, we cannot disregard the potential for overfitting, prompting the incorporation of dropout mechanisms. The architecture can be represented as follows:
%It is worth noticing, \citep{buchel2022deep} employs millions of swaptions trained over 5 000 epochs, \citep{liu2019neural} uses 8 000 epochs for one million data points. 
%\vfill \mbox{}
% NEURAL NETWORK
\vspace{-0.4cm}
\begin{figure}[H]
    \centering
    \begin{tikzpicture}[x=3.05cm,y=1.cm]
      \readlist\Nnod{3,4,6,4,2} % array of number of nodes per layer
      \readlist\Nstr{nf, h1, h2, h3, np} % array of string number of nodes per layer
      \readlist\Cstr{u,h^{(\prev)},v} % array of coefficient symbol per layer
      \def\yshift{0.55} % shift last node for dots
      
      % LOOP over LAYERS
      \foreachitem \N \in \Nnod{
        \def\lay{\Ncnt} % alias of index of current layer
        \pgfmathsetmacro\prev{int(\Ncnt-1)} % number of previous layer
        \foreach \i [evaluate={\c=int(\i==\N); 
                               \y=\N/2-\i-\c*\yshift;
                               \x=\lay; 
                               \n=\nstyle;
                               \index=(\i<\N?int(\i):"\Nstr[\Ncnt]");}] in {1,...,\N}{ % loop over nodes
                               %\index="\Nstr[\N]";}] in {1,...,\N}{ % loop over nodes
          % NODES
          \node[node \n] (N\lay-\i) at (\x,\y) {$\strut\Cstr[\n]_{\index}$};
          
          % CONNECTIONS
          \ifnumcomp{\lay}{>}{1}{ % connect to previous layer
            \foreach \j in {1,...,\Nnod[\prev]}{ % loop over nodes in previous layer
              \draw[-stealth] (N\prev-\j) -- (N\lay-\i);
              \draw[-stealth] (N\prev-\j) -- (N\lay-\i);
            }
            \ifnum \lay=\Nnodlen
              \draw[-stealth] (N\lay-\i) --++ (0.5,0); % arrows out
            \fi
          }
          {
            \draw[-stealth] (0.5,\y) -- (N\lay-\i); % arrows in
          }
        }
        \path (N\lay-\N) --++ (0,1+\yshift) node[midway,scale=1.6] {$\vdots$}; % dots
      }
      
      % LABELS
      \node[above=3,align=center,mydarkgreen] at (N1-1.90) {Input\\[-0.2em]layer};
      \node[above=2,align=center,mydarkblue] at (N2-1.90) {Hidden\\[-0.2em]layers};
      \node[above=2,align=center,mydarkblue] at (N4-1.90) {Has dropout (0.25)};
      \node[above=3,align=center,mydarkred] at (N\Nnodlen-1.90) {Output/Prediction \\[-0.2em]layer};
    \end{tikzpicture}

    \vspace{-0.4cm}
    \caption{FCN Architecture of the Indirect Deep Calibration}
\end{figure}

Additionally, we use mini-batches of size 1\,000 and train for 1\,000 epochs. The Adam algorithm \citep{kingma2014adam} is used for optimization, without weight decay, and with a learning rate of 0.001. The error to minimize is the MSE (see Equation \eqref{eq:MSE}).

\subsection{Direct Deep Calibration of the G2++ model}
\label{sec:DirectDC}
\subsubsection{ZCs as OQOI}\mbox{}\\
\label{sec:zcasoqoi}
The aim of this subsection is to consider a calibration process that directly uses ZC rate curves\footnote{We could have similarly used FWD rates, but the idea of the Direct DC is to reduce data transformations as much as possible. It is worth noting that ZC rates are directly observed in the market, which is not necessarily the case for FWD rates.} as they are direct market observations, unlike correlations and covariances. This approach makes the process easier to use and reduces computational costs since no further transformation is needed. Indeed, ZC rates can be directly plugged into our model. For calibrations, we generally want to capture historical behavior. To do so, we consider simulations of ZC rates on several dates and maturities. This means we have to deal with matrices, where the number of rows corresponds to the temporal depth, and the number of columns corresponds to the number of maturities. Due to the nature of these inputs, it seems that for this purpose, a CNN is better suited than an FCN.

\subsubsection{Data set construction}\mbox{}\\
\label{sec:dataZC}
The process is mostly the same as that for indirect DC. Only step 3 (actual computation of the OQOI) changes\footnote{Regarding the scaling transformations, using the min-max scaler only on the targets gave better results.}.

\vspace{0.5cm}
\paragraph{\textbf{Computing ZC rates:}}\mbox{}\\ 
From equations \eqref{eq:Z} and \eqref{eq:G2pp}, taking $x_0 = y_0 = 0$, we obtain an explicit formula for the expectation of ZC rates given by:
\begin{equation}
    \label{eq:expecZCs}
    \mathbb{E}\left[Z(t, T) \right] = -\frac{1}{T-t}\log \left[\frac{P^{M}(0, T)}{P^{M}(0, t)} \right] -\frac{1}{2(T-t)} \left[V(t, T)-V(0, T)+V(0, t)\right].
\end{equation}
Taking the expectation of ZC rates as the input for our direct DC model allows us to obtain a typical path that could be observed in the market (see Figure \ref{fig:mkt0}). Note that in practice, real market ZC rate curves will be used as input when calibrating with our model.

For the initial market bond price, $P^M(0,t)$, we take the Euro curve as of 2020/11/04, and $V(.,.)$ is explicitly defined in \eqref{eq:VV}. Unlike the indirect calibration using an FCN, with a CNN we do not have issues with the number of tenors. As stated before, when considering the initial market curve, we can observe an inhomogeneous behavior in the short-term tenors compared to the long-term ones. Taking enough maturities into account significantly reduces the influence of this behavior. Hence, we take the following sets of maturities: $\{1\;\text{day}, \;1\;\text{week}, \;1\;\text{month}, \;2\;\text{months}, \;3\;\text{months}, \;6\;\text{months}, \;9\;\text{months}\}$; $[1\;\text{year}, \;12\;\text{years}]$ with a 1-year step; and $[15\;\text{years}, \;50\;\text{years}]$ with a 5-year step. Finally, for the propagation, expectations of ZCs will be computed for time steps $t_i = i\Delta, \; i \in [|1, nb_{\text{steps}}|]$, with the step size $\Delta = \text{1\;week}$, and we make 105 steps of propagation (about 2 years). Thus, our inputs will be in $\mathbb{R}^{nb_{\text{steps}} \times nb_{\text{tenors}}}$ with $nb_{\text{steps}} = 106$ and $nb_{\text{tenors}} = 28$.

\begin{figure}[H]
    \centering
    \includegraphics[height=6.5cm]{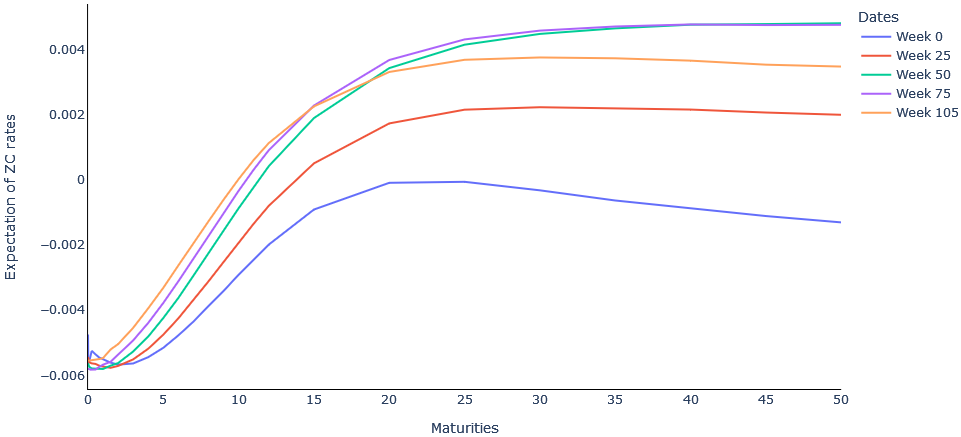} % first figure itself
    \caption{Example of expectations of simulated ZC rate curves for different dates. $K_x=0.07$, $K_y=0.09$, $\sigma_x=0.09$, $\sigma_y=0.09$ and $\rho=-0.99$.}
    \label{fig:zcgen}
\end{figure}

Figure \ref{fig:zcgen} shows, for a randomly selected set of parameters, the generated ZC rate curves for a set of projection dates. Appendix \ref{sec:apdx_ZC_gen} shows actual market ZC rates. We can observe similar curves.

\subsubsection{The neural network architecture for direct DC}\mbox{}\\
\label{sec:archiDDC}
The architecture combining convolution and linear layers is still quite shallow. We start with a convolution layer of dimension (C=1, H=106, $nb_{\text{tenors}} = 28$), with a filter of size $(7 \times 7)$. The stride of the convolution is 2, and we also allow padding. Then, we add a pooling layer with a stride of 2. Finally, two linear layers follow, the first with 100 neurons and the second one (which is the prediction layer) with $np=5$ neurons. Here as well, we include a dropout with a probability of 0.25 in the second-to-last layer.

%\newpage
\begin{figure}[H]
	\centering
	\begin{tikzpicture}
		\node at (0.5,-1){\begin{tabular}{c}ZCs $\in \mathbb{R}^{nb_{steps} \times nb_{tenors}}$\\layer $l = 0$\end{tabular}};
		
		\draw (0,0.5) -- (1,0.5) -- (1,1.5) -- (0,1.5) -- (0,0.5);
		
		\draw[-stealth] (1,0.9) -- (2+1.5,0.9);
		\draw[-stealth] (1,1.) -- (2+1.5,1); % arrows in
		\draw[-stealth] (1,1.1) -- (2+1.5,1.1);
		
		\node at (2.5+1.5,3.5){\begin{tabular}{c}Convolutional layer\\with 7 x 7 filter\\layer $l = 1$\end{tabular}};
		
		\draw[fill=black,opacity=0.2,draw=black] (2+1.5,0.5) -- (3+1.5,0.5) -- (3+1.5,1.5) -- (2+1.5,1.5) -- (2+1.5,0.5);
		\draw[fill=black,opacity=0.2,draw=black] (2.5+1.5,1.25) -- (3.25+1.5,1.25) -- (3.25+1.5,2) -- (2.5+1.5,2) -- (2.5+1.5,2);

		\draw[-stealth] (3.25+1.5,0.9) -- (4.25+2.25,0.9);
		\draw[-stealth] (3.25+1.5,1.) -- (4.25+2.25,1); % arrows in
		\draw[-stealth] (3.25+1.5,1.1) -- (4.25+2.25,1.1);
		
		\node at (4.75+2.25,-1){\begin{tabular}{c}Pooling (Max) layer\\Kernel size = 7\end{tabular}};
		
		\draw[fill=black,opacity=0.2,draw=black] (4.25+2.25,0.5) -- (5+2.25,0.5) -- (5+2.25,1.25) -- (4.25+2.25,1.25) -- (4.25+2.25,0.5);
		
		\draw[-stealth] (5+2.25,0.9) -- (10.5-0.3,1.70);
		\draw[-stealth] (5+2.25,1.1) -- (10.5-0.8,1.1);
		\draw[-stealth] (5+2.25,0.7) -- (10.5-1.2,0);
		\draw[-stealth] (5+2.25,0.7) -- (10.5-1.2,0.4);

		\node at (12-1.75,3.5){\begin{tabular}{c}Fully connected layer\\100 neurons and dropout(0.25)\end{tabular}};
		
		\draw[fill=black,draw=black,opacity=0.5] (10.5-1.75,0) -- (11-1.75,0) -- (12.5-1.75,1.75) -- (12-1.75,1.75) -- (10.5-1.75,0);
		
		\draw[-stealth] (10.5-0.5,0.9) -- (13.25,0.7);
		\draw[-stealth] (10.5-0.67,1.1) -- (13.0,1.1);
		\draw[-stealth] (10.5-1.55,0.2) -- (13.25,0.85);
		\draw[-stealth] (10.5-0.5,1.4) -- (12.85,0.5);
		
		\node at (12,-1){\begin{tabular}{c}Fully connected \\ output /  prediction \\ layer \\ 5 neurons\end{tabular}};
		
		\draw[fill=black,draw=black,opacity=0.5] (12.5,0.5) -- (13,0.5) -- (13.65,1.25) -- (13.15,1.25) -- (12.5,0.5);
	\end{tikzpicture}
    \vspace{-0.7cm}
	\caption{CNN Architecture of the Direct Deep Calibration}
	\label{fig:traditional-convolutional-network}
\end{figure}

Similarly to the indirect DC, we use mini-batches of size 1,000, and the Adam algorithm without weight decay is applied to minimize the MSE. The learning rate is 0.0002, and the number of epochs is 4,000.

\section{Results for indirect and direct DC models}
\label{sec:resec}
Regarding accuracy, both algorithms produce good results. Time efficiency is also remarkable, with calibration on the test set (2,000 entries) performed in less than 0.30 seconds for the indirect DC model and less than 0.10 seconds for the direct DC model. Table \ref{table:res} below summarizes the MSEs obtained on the $n_{test}$ observations of the test set for each parameter, given by:
\begin{equation}
    \text{MSE}(\theta^i, \hat{\theta}^i) = \frac{1}{n_{test}}\sum_{j=1}^{n_{test}}(\theta^i_j-\hat{\theta}^i_j)^2, \text{ for i $\in$ $\llbracket 1,\;np \rrbracket$}.
    \label{eq:MSEform1}
\end{equation}
\begin{table}[H]
    \centering % used for centering table
    \begin{tabular}{c | c || c c c c c} % centered columns (5 columns)
        \hline \hline  % inserts double horizontal line
        Method & OQOI & $K_x$ & $K_y$ & $\sigma_x$ & $\sigma_y$  & $\rho$ \\ [1.ex] % inserts table
        \hline \hline %inserts double horizontal lines
         & & & & \\
          & COV-ZC	& $3.5 \times 10^{-4}$	& $5.6 \times 10^{-4}$	& $7.7 \times 10^{-4}$	& $8.3 \times 10^{-4}$	& $8.2 \times 10^{-2}$\\ [1.ex] % inserting body of the table
        \multirow{2}{4em}{Indirect DC} & COV-FWD	& $3.5 \times 10^{-4}$	& $5.5 \times 10^{-4}$	& $7.6 \times 10^{-4}$	& $8.3 \times 10^{-4}$	& $8.2 \times 10^{-2}$\\ [1.ex] % inserting body of the
          & COR-ZC	& \color{black}$7.5 \times 10^{-4}$	& \color{black}$1.1 \times 10^{-3}$	& \color{black}$1.3 \times 10^{-3}$	& \color{black}$1.4 \times 10^{-3}$	& \color{black}$2.0 \times 10^{-1}$\\ [1.ex] % inserting body of the table
          & COR-FWD	& \color{black}$7.4 \times 10^{-4}$	& \color{black}$1.1 \times 10^{-3}$	& \color{black}$1.3 \times 10^{-2}$	& \color{black}$1.4 \times 10^{-2}$	& \color{black}$2.4 \times 10^{-2}$\\ [1.ex] % inserting body of the table
        \hline \hline 
        Direct DC & ZCs	& $4.0 \times 10^{-4}$	& $6.3 \times 10^{-4}$	& $9.8 \times 10^{-4}$	& $1.0 \times 10^{-4}$	& $1.52 \times 10^{-1}$\\ [1.ex] % inserting body of the
        \hline \hline 
    \end{tabular}
    \vspace{1.5mm}
    \caption{Calibration errors on the test set} % title of Table
    \label{table:res} % is used to refer this table in the text
\end{table}

\begin{figure}[H]
    \centering
    \begin{minipage}{0.5\textwidth}
        \begin{figure}[H]
            \centering
            \includegraphics[width=9.25cm, height=6cm]{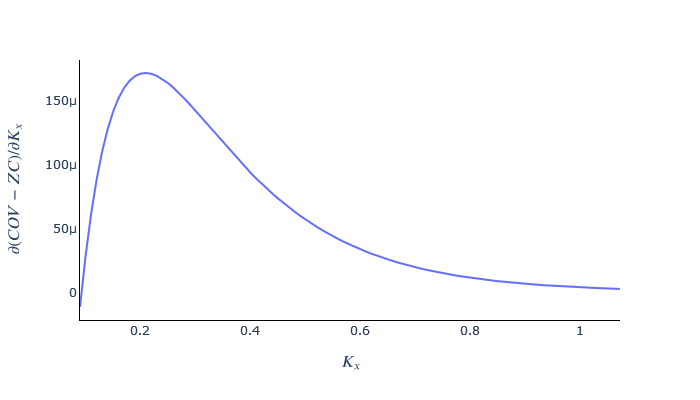} % first figure itself
            \vspace{-9.5mm}
            \caption{Derivative of COV-ZC (5Y, 7Y) w.r.t. $K_x$}
            \label{fig:der1}
        \end{figure}
    \end{minipage}\hfill
    \begin{minipage}{0.5\textwidth}
        \begin{figure}[H]
            \centering
            \includegraphics[width=9.25cm, height=6cm]{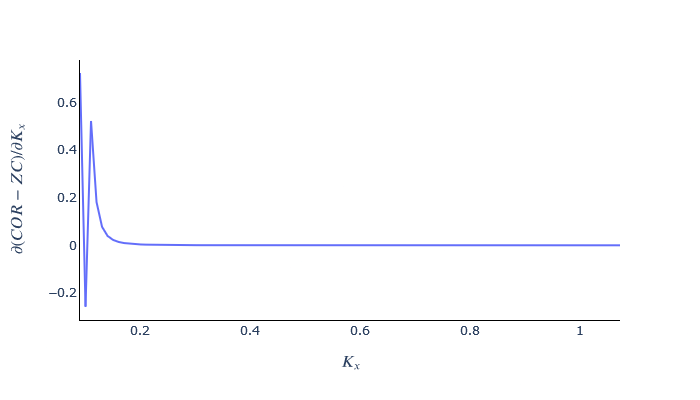} % second figure itself
            \vspace{-9.5mm}
            \caption{Derivative of COR-ZC (5Y, 7Y) w.r.t. $K_x$}
            \label{fig:der2}
        \end{figure}
    \end{minipage}
\end{figure}

\begin{figure}[H]
    \centering
    \begin{minipage}{0.5\textwidth}
        \begin{figure}[H]
            \centering
            \includegraphics[width=9.25cm, height=6cm]{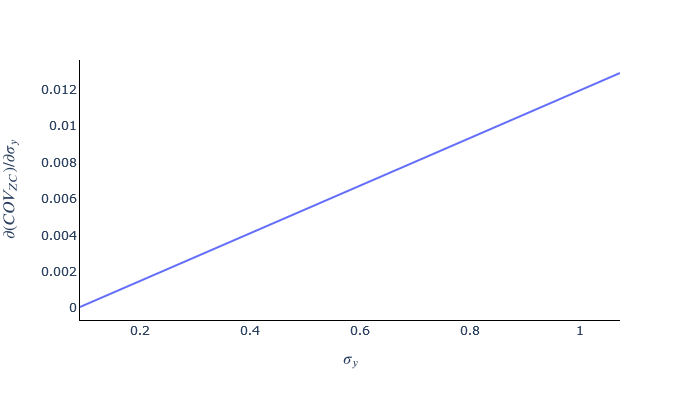} % first figure itself
            \vspace{-9.5mm}
            \caption{Derivative of COV-ZC (5Y, 7Y) w.r.t. $\sigma_y$}
            \label{fig:der3}
        \end{figure}
    \end{minipage}\hfill
    \begin{minipage}{0.5\textwidth}
        \begin{figure}[H]
            \centering
            \includegraphics[width=9.25cm, height=6cm]{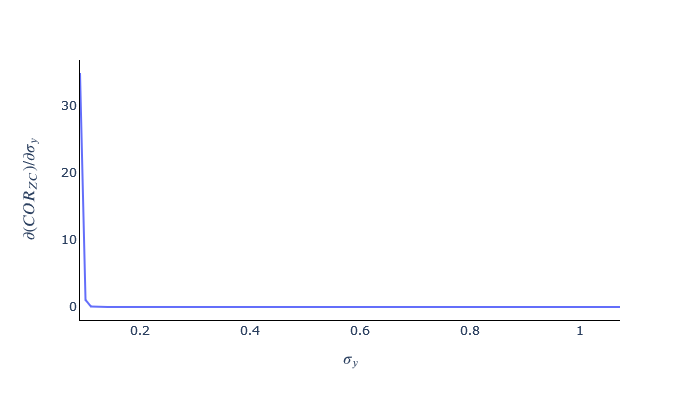} % second figure itself
            \vspace{-9.5mm}
            \caption{Derivative of COR-ZC (5Y, 7Y) w.r.t. $\sigma_y$}
            \label{fig:der4}
        \end{figure}
    \end{minipage}
\end{figure}

\begin{figure}[H]
    \centering
    \begin{minipage}{0.5\textwidth}
        \begin{figure}[H]
            \centering
            \includegraphics[width=9.25cm, height=6cm]{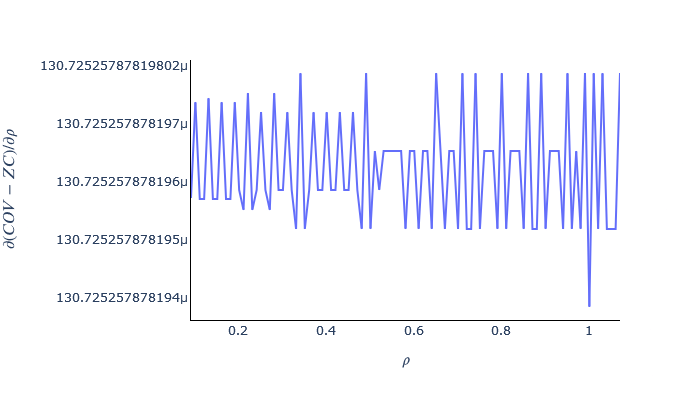} % first figure itself
            \vspace{-9.5mm}
            \caption{Derivative of COV-ZC (5Y, 7Y) w.r.t. $\rho$}
            \label{fig:der5}
        \end{figure}
    \end{minipage}\hfill
    \begin{minipage}{0.5\textwidth}
        \begin{figure}[H]
            \centering
            \includegraphics[width=9.25cm, height=6cm]{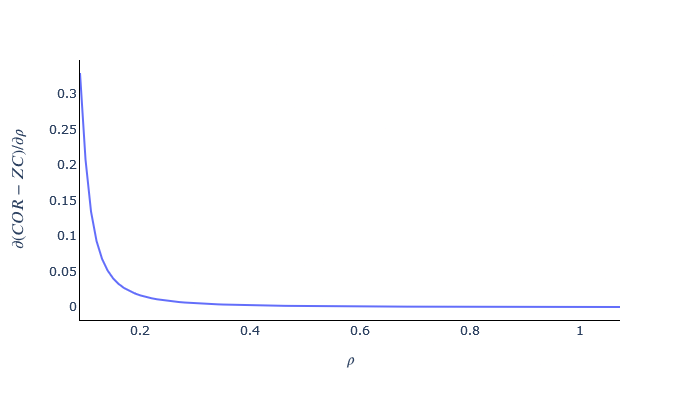} % second figure itself
            \vspace{-9.5mm}
            \caption{Derivative of COR-ZC (5Y, 7Y) w.r.t. $\rho$}
            \label{fig:der6}
        \end{figure}
    \end{minipage}
\end{figure}

\paragraph{\textbf{Indirect DC}}\mbox{}\\ 
As one might expect, given the similarities between the FWD and ZC rate curves, the results with the indirect DC are very close for FWDs and ZCs rates. The most interesting result we observe is that the error using covariances is about half the error using correlations. It appears that covariances provide significantly more information for the calibration process. We can verify this numerically by observing the derivative curves of covariances and correlations with respect to the parameters, computed by finite differences. We arbitrarily select the maturity pair (5Y, 7Y) and a subset of 100 parameters for each of $K_x, \, \sigma_y,$ and $\rho$, and compute the derivatives. Figures \ref{fig:der1}, \ref{fig:der3}, and \ref{fig:der5} show the derivatives of ZC rates covariances for the sample of parameters, while Figures \ref{fig:der2}, \ref{fig:der4}, and \ref{fig:der6} show the derivatives of ZC rates correlations for the same sample. We can directly see that the correlation derivatives quickly vanish for the three parameters. Meanwhile, for covariances, only the derivative for $K_x$ vanishes, and it does so less quickly. For $\sigma_y$ and $\rho$, we do not observe this vanishing behavior at all. Similar results are obtained when observing the other parameters $K_y$ and $\sigma_x$.

\vspace{-5mm}

\vspace{-12.5mm}

\vspace{-12.5mm}

Thanks to this analysis, we observe a phenomenon with effects similar to the vanishing gradient. When this phenomenon occurs, it prevents an effective learning process using backpropagation on correlations. In fact, we present a theorem explaining when backpropagation becomes unfeasible. To our knowledge, such an intuitive result does not appear in the literature.

\begin{theorem}[Unfeasible backpropagation]
\label{theo:unfeasbackprop}
In a feedforward neural network architecture, if the derivatives of the inputs with respect to the targets vanish, then the derivatives of the loss function with respect to the weights also vanish.
\end{theorem}

The proof of this theorem is provided in Appendix \ref{sec:proof}. A straightforward consequence of this theorem is that when the derivatives of the loss function with respect to the weights vanish, it prevents gradient descent by backpropagation, subsequently stalling the training process. Our indirect DC provides a practical illustration of the theorem. The derivatives of the correlation with respect to the parameters (which are the targets of our NN architecture) vanish, as shown in Figures \ref{fig:der2}, \ref{fig:der4}, and \ref{fig:der6}. Consequently, the gradient descent will stall. Thus, this theorem clarifies why covariances are more suitable than correlations for our DC purposes. More generally, it provides practitioners with a rule of thumb to avoid some cases of stalling gradients. 

\vspace{0.5cm}
\paragraph{\textbf{Direct DC}}\mbox{}\\
Compared to the indirect DC with ZC rates covariances, the direct method leads to slightly less accurate calibrations. Although the differences are not significant, this outcome makes sense. The indirect DC benefits from information already extracted from the ZC rate curve, taking advantage of the deeper insights provided by the covariance matrices.

\begin{comment}
Improving this method would be of great use as it is for the practitioner, the easiest to use. The first idea would be to have a deeper reflection on the tenors that really bring information, choosing them more wisely could lead to better results and better computational performances. Another lead could be to challenge the CNN architecture, as the data is more complex than indirect DC's inputs, choosing the right pooling and activation functions, the right hyper-parameters (including the number of layers, i-e the depth of the CNN) and so on, could have a great impact on the accuracy. 
\end{comment}

\begin{figure}[H]
    \centering
    \includegraphics[width=15.5cm, height=4.5cm]{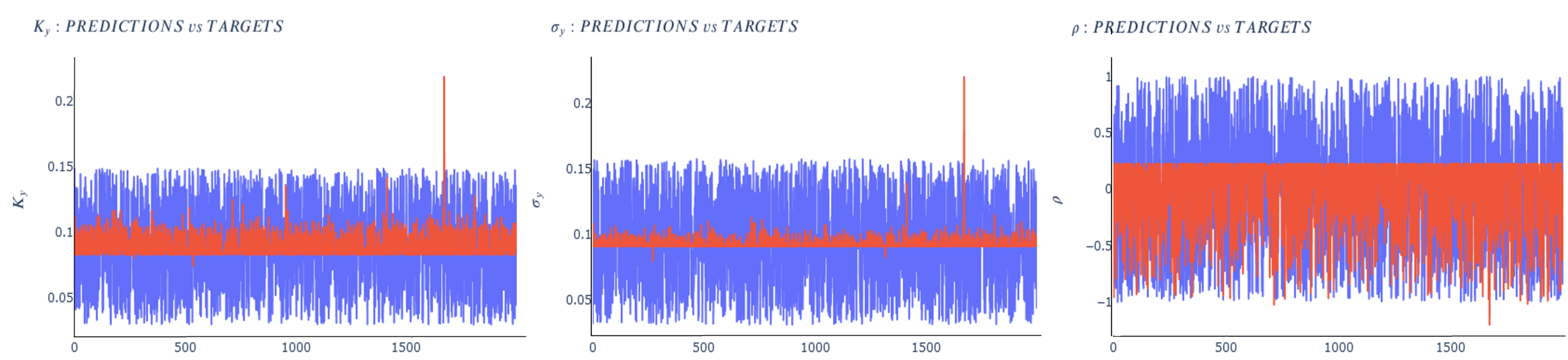} % second figure itself
    \vspace{-0.7cm}
    \caption{Indirect DC on FWD rates correlations: fitting curve on $K_y,\; \sigma_y, \text{ and } \rho$. The red curves are predictions and the blue curves actual values.}
    \label{fig:cordwd_calib}
\end{figure}

\begin{figure}[H]
    \centering
    \includegraphics[width=15.5cm, height=7.5cm]{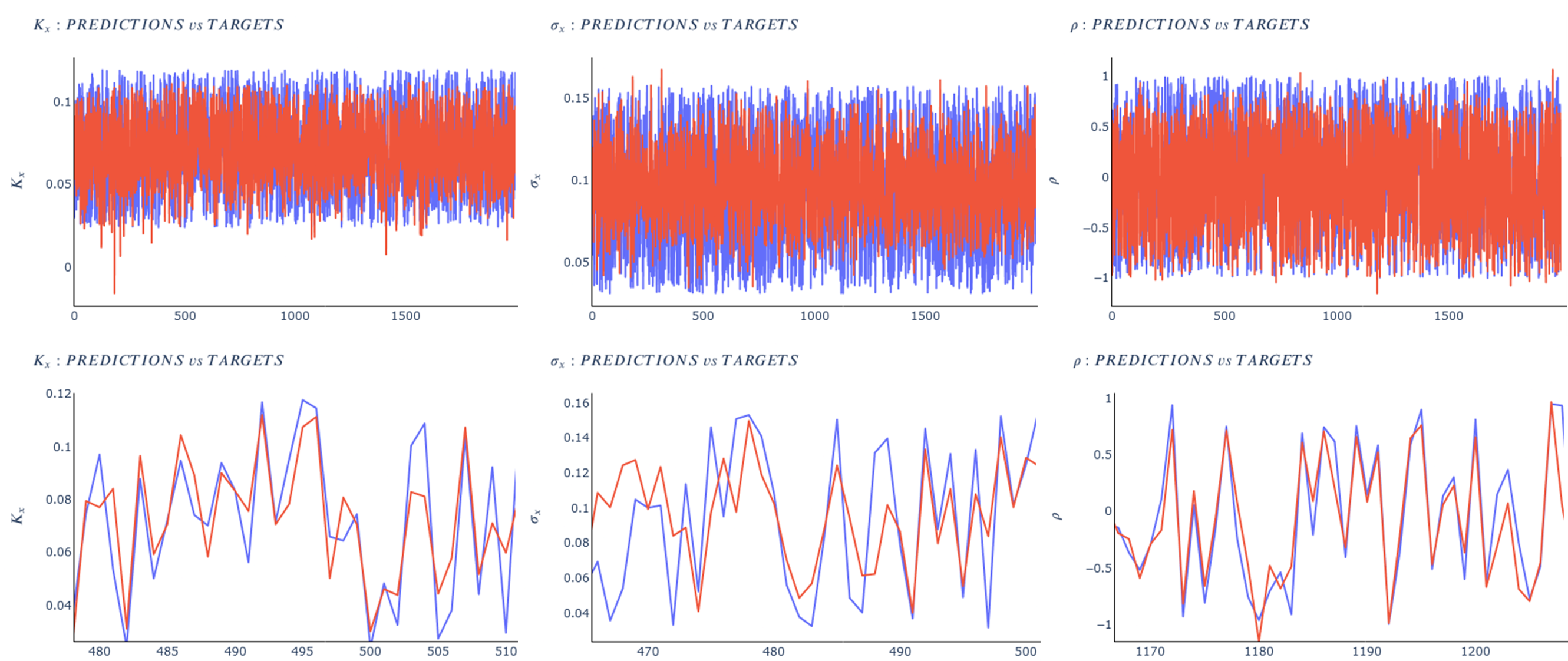} % second figure itself
    \vspace{-0.7cm}
    \caption{Indirect DC on ZC rates covariances: fitting curve on $K_x,\; \sigma_x, \text{ and } \rho$ (original and zoomed-in). The red curves are predictions and the blue curves actual values.}
    \label{fig:covzc_calib}
\end{figure}

\begin{figure}[H]
    \centering
    \includegraphics[width=15.5cm, height=7.5cm]{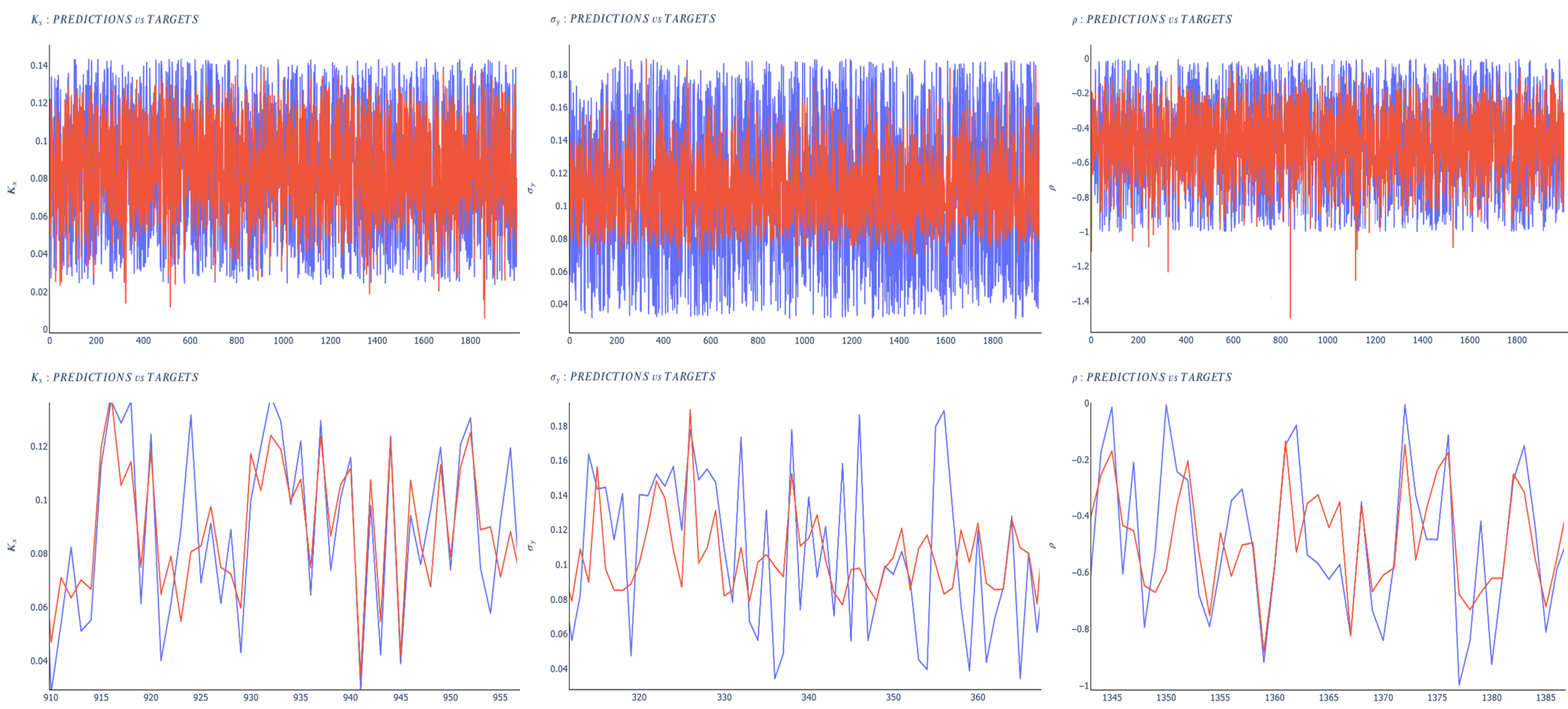} % second figure itself
    \vspace{-0.7cm}
    \caption{Direct DC on ZC rates: fitting curve on $K_x,\; \sigma_x, \text{ and } \rho$ (original and zoomed-in). The red curves are predictions and the blue curves actual values.}
    \label{fig:figZC}
\end{figure}

\vspace{0.5cm}
\paragraph{\textbf{Fitting curves}}\mbox{}\\ 
While MSE analysis allows for the comparison of different models, visualizations enable the observation of the fitting behavior. Figure \ref{fig:covzc_calib} shows the fitting on the test dataset for the indirect DC with ZC rates covariances, including a zoomed-in view on a randomly selected interval, and Figure \ref{fig:cordwd_calib} shows the fitting for the correlations. Finally, Figure \ref{fig:figZC} shows the fitting for the direct DC with ZC rates.

\vspace{-0.35cm}


These visualizations further support the fact that covariances are better than correlations for indirect DC and that direct DC's performance is close to that of indirect DC. It is worth noting that, from a practical point of view, using direct DC, which works on direct market observations (unlike covariances and correlations), is easier to use and has a reduced computational cost since no further transformation is needed.

\section{Deep Calibration vs Classic Calibration}
\label{sec:compsec}

\subsection{Comparison Methodology}\mbox{}\\
\label{sec:compmethodo}
%\begin{comment}
It is not a trivial task to perform a comparison between our DC models and classic optimization methods for calibration. Assume that we want to compare a model using swaptions to perform a calibration with our DC model (regardless of whether the DC is indirect or direct). We would obtain one set of parameters from those swaptions and another set of parameters from our calibration using ZC or FWD rates (directly or indirectly). There is no guarantee that the two sets of parameters will be alike since we are not trying to fit the same quantities, even if the calibration were perfect in both cases. This similarly happens when choosing between swaptions or caplets to calibrate a classic model. Consequently, it is impractical to use data not related to our OQOIs. Therefore, we focus on calibrations based on ZC and FWD rates and their correlations and covariances.
%\end{comment}

A common calibration method widely used in practice consists of finding the set of parameters that minimizes the error between the theoretical covariances/correlations of ZC and FWD rates obtained from equation \eqref{eq:covcor2} and the historical ones. We choose this method as the benchmark classical calibration (CC) method to compare against our DC method. To do so, we use a synthetic set of parameters. Indeed, using real market data would make it difficult to compare the efficiency of both calibration methods since the real set of parameters targeted would be unknown. Below, we outline the different steps of our comparison process between DC and CC.

\begin{enumerate}
    \item First, we choose $N_{sets}$ sets of parameters $\left\{k_{x}^i, k_{y}^i, \sigma_{x}^i, \sigma_{y}^i, \rho^i\right\}_{i \in\left[ \left|1, N_{sets}\right| \right]}$ for the comparison.
    \item We compute ZCs with equation \eqref{eq:Z} and add white noise and jumps to replicate real market data behavior, including stress periods.
    \item Next, we compute (numerically) the covariances and correlations from the obtained ZC rates, which will only be useful for the CC.
    \item We proceed with the direct DC and CC.
    \item Finally, we compare the two calibrations.
\end{enumerate}

To measure how sensitive both methods are to the noise added in step two above and how they perform in a non-stressed environment, we will also make the comparison with the data without the noise. It is worth noting that in this process, we chose to use the direct DC instead of the indirect DC for two main reasons:

\begin{itemize}
    \item We have already compared the direct DC and the indirect DC in the previous section. The latter showed very similar calibration performances, and as we will see in Subsection \ref{sec:dcvscc}, the direct DC performs better than the CC.
    \item As explained in Section \ref{sec:zcasoqoi}, the direct DC is clearly advantageous not only in terms of computational cost compared to the indirect DC but also in terms of practical use. Hence, the choice of performing the comparison using the direct DC, which would be the natural first choice for any practical user.
\end{itemize}

\vspace{0.5cm}
\paragraph{\textbf{Step 1: Choice of the set of parameters}}\mbox{}\\
To make a fair comparison, the parameters must be outside the training dataset and ideally even the test set. We keep the same range of parameters defined in Table \ref{table:rangeparams} and uniformly draw $N_{sets}$ sets of parameters. Setting $N_{sets} = 50$ would be a reasonable choice, especially from a computational point of view, and would make sense in a real market situation. It could correspond to a scenario where a market risk unit needs to perform G2++ calibrations for ten currencies in five different stress scenarios.

\vspace{0.5cm}
\paragraph{\textbf{Step 2: Adding noise to ZC rate curves to mimic real market stressed data}}\mbox{} \\ 
We compute the expectation of ZC rates (see equation \eqref{eq:expecZCs}) at times $t_i = i \Delta, \, i \in [|1, nb_{steps}|]$ using the results in Subsection \ref{sec:dataZC}. To approach real-world data, we decide to add two types of noise. First, we add jumps, and then we add Gaussian white noise. For a given tenor $T$, we denote $\Tilde{Z_i}$ as the ZC rate at time $t_i$ with added jumps and $\hat{Z_i}$ as the ZC rate with both jumps and Gaussian noise. Assuming positive ZC rates, the process is as follows:

\begin{itemize}
    \item Following a Bernoulli distribution with parameter $p$, we decide whether a jump will occur at time $t_i$ (we take $p = 2\%$). If a jump occurs, it will have a magnitude $\alpha_i$, which is a uniform random variable in $\left[\frac{1}{a}, \;a\right]$.
    
    \item If a jump occurs at time $t_i$, there will be no jump at times $t_{i+1}$ and $t_{i+2}$. Instead, we will have a correction with $\Tilde{Z}_{i+1} = \beta_i Z_{i}$ and $\Tilde{Z}_{i+2} = \gamma_i Z_{i}$, where $\beta_i$ is uniformly drawn between $\alpha_i$ and $1$. Similarly, $\gamma_i$ is drawn from a uniform distribution between $\beta_i$ and $1$.
\end{itemize}

\begin{figure}[H]
    \centering
    \includegraphics[height=10cm, width= 13cm]{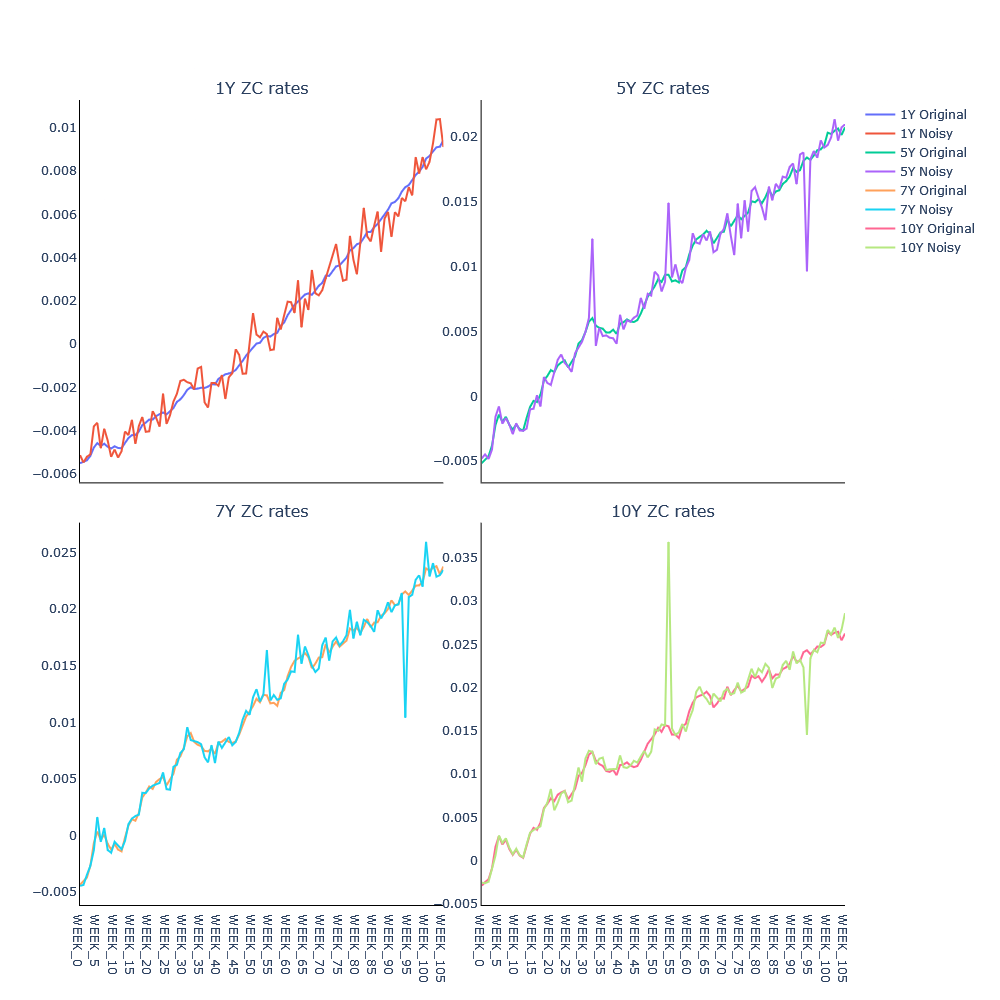} 
    % \vspace{-0.5cm}
    \caption{Stressed and not stressed ZC rate curves. $K_x=0.09$, $K_y=0.08$, $\sigma_x=0.13$, $\sigma_y=0.06$ and $\rho=-0.96$.}
    \label{fig:noisy3}
\end{figure}

The first step is motivated by the fact that we know jumps occur in the market, but they are not \textit{extremely} frequent. The second step allows us to better reflect market behavior, where jumps are generally not followed by an immediate return to normality but often require some time before stabilizing. The process described above, in the case of a jump, can be written as:
\begin{equation}
    \begin{cases}
        & \Tilde{Z}_i = \alpha_i \times Z_{i} \; with \; \alpha_i \sim U_{\left[\frac{1}{a},\;a\right]} \\
            & \Tilde{Z}_{i+1} = \beta_i \times Z_{i} \; with \; \beta_i \sim U_{[min(1, \;\alpha_i),\;max(1, \;\alpha_i)]} \\
        & \Tilde{Z}_{i+2} = \gamma_i \times Z_{i} \; with \; \gamma_i \sim U_{[min(1, \;\beta_i),\;max(1, \;\beta_i)]}. \\
    \end{cases} 
\end{equation}
We chose $a=\frac{3}{2}$. To take different tenors into consideration, one needs to use different intensities for the jumps. To do so, we define the sets $\left\{\alpha_i^{0}, \; \dots, \;\alpha_i^{M-1}\right\}, \, \left\{\beta_i^{0}, \; \dots, \;\beta_i^{M-1}\right\},$ and $\left\{\gamma_i^{0}, \; \dots, \;\gamma_i^{M-1}\right\}$, where $M$ defines the number of tenors (typically, based on the previous notations when describing the NN architecture for the DC in Subsection \ref{sec:archiDDC}, $M = nb_{tenors}$). We then apply the previous process to the first tenor with $\alpha_i^0 = \alpha_i$, $\beta_i^0 = \beta_i$, and $\gamma_i^0 = \gamma_i$, and subsequently get the coefficients for $j \in \llbracket 1,\,M-1 \rrbracket$ using the following process.

\begin{equation}
    \begin{cases}
        & \alpha_i^j = \max(1,\,\alpha_i^0\;U_i^j) \text{ if } \alpha_i^0 >= 1 \text{ and } \alpha_i^j = \min(1,\,\alpha_i^0\;U_i^j) \text{ otherwise }\\
        & \beta_i^j = U_{[min(1, \;\alpha_i^j),\;max(1, \;\alpha_i^j)]} \\
        & \gamma_i^j = U_{[min(1, \;\beta_i^j),\;max(1, \;\beta_i^j)]}, \\
    \end{cases} 
\end{equation}
with $U_i^j$ drawn for each $j$ such that, $U_i^j \sim U_{\left[b,\;c\right]}$.
We choose $b=\frac{1}{2}$ and $c=\frac{3}{2}$. This leads us to avoid having jumps of same magnitude for all tenors while keeping same signs (increasing or decreasing stress). For negative ZC rates, $Z_i$, at any considered time step $t_i$ and for a given tenor, the related coefficients, $\alpha_j,\,\beta_j, \text{ or } \gamma_j$ is inverted. Regarding the Gaussian noise, we will add $M$ centered Gaussian random variables with a standard deviation proportional to the one observed on the ZC rate curve at time $t_i$, (i-e, considering the whole term structure) , we denote it $\sigma_{Z_i}$ . For the ZC rate of tenor $j$ at time $t_i$, we hence write $\hat{Z}_i^j = \Tilde{Z}_i^j + N(0,\,(d\sigma_{Z_i})^2)
$. We choose $d=10\%$. Figure \ref{fig:noisy3} shows for one set of parameters  original and noisy ZCs curves. 

We choose $b=\frac{1}{2}$ and $c=\frac{3}{2}$. This helps avoid having jumps of the same magnitude for all tenors while maintaining the same signs (increasing or decreasing stress). For negative ZC rates, $Z_i$, at any considered time step $t_i$ and for a given tenor, the related coefficients, $\alpha_j$, $\beta_j$, or $\gamma_j$, are inverted. Regarding the Gaussian noise, we will add $M$ centered Gaussian random variables with a standard deviation proportional to that observed on the ZC rate curve at time $t_i$ (i.e., considering the whole term structure). We denote it $\sigma_{Z_i}$. For the ZC rate of tenor $j$ at time $t_i$, we thus write $\hat{Z}_i^j = \Tilde{Z}_i^j + N(0,\,(d\sigma_{Z_i})^2)$. We choose $d=10\%$. Figure \ref{fig:noisy3} shows the original and noisy ZC curves for one set of parameters.

%\newpage

%\vspace{-.7cm}

\begin{figure}[H]
    \centering
    \includegraphics[width=16.5cm, height=6.75cm]{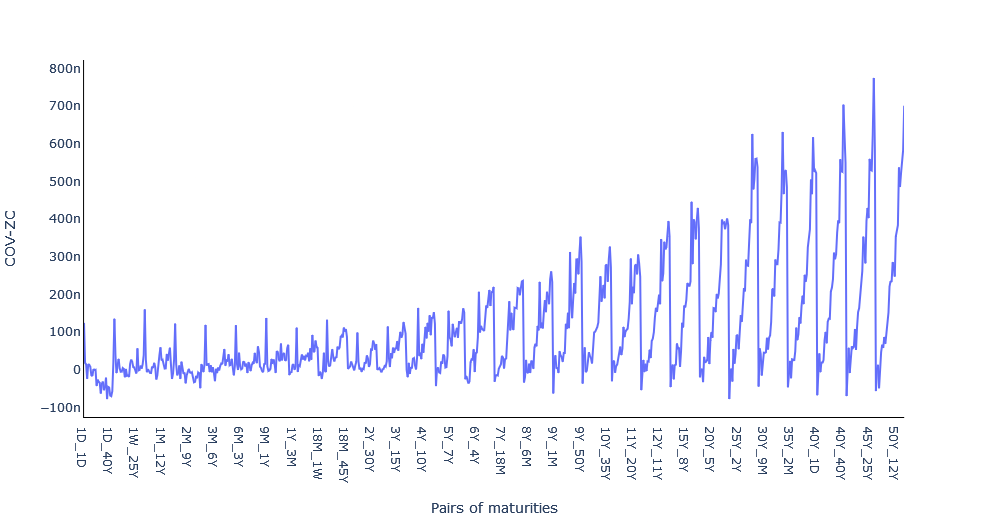} % second figure itself
    \caption{Covariance curve of ZC rates after add of noise. $K_x=0.07$, $K_y=0.07$, $\sigma_x=0.03$, $\sigma_y=0.03$ and $\rho=-0.42$.}
    \label{fig:covnoisy}
\end{figure}

\paragraph{\textbf{Step 3: Numerical computation of COVs and CORs}}\mbox{}\\
As Figure \ref{fig:covnoisy} shows (on covariances), the noise added to ZC rates has an impact on correlations and covariances as they also become less smooth. 

\paragraph{\textbf{Step 4: CC and DC}} \mbox{}\\ 
For CC, the purpose is to find the set of parameters that minimizes the error between real correlations and covariances and their estimates via equation \eqref{eq:covcor1}. We proceed with the following minimization on each of our $N_{sets} = 50$ sets of parameters. Let us denote $\theta_i = (K_x^i, K_y^i, \sigma_x^i, \sigma_y^i, \rho^i)$, where $K_x^i, K_y^i, \sigma_x^i, \sigma_y^i > 0$, and $\rho^i \in [-1,1]$, with $i \in \left[\left|1, N_{sets}\right|\right]$. Then, 
\begin{equation}
    \begin{aligned}
        \theta_i^{*} = \underset{\theta_i}{\mathrm{argmin}} \frac{1}{M}\sum_{i,j=1}^M & \left[W_{cov} \left|\frac{{\rm Cov}(T_i,T_j;\,\theta_i) - {\rm Cov}(T_i,T_j)^{noisy}}{{\rm Cov}(T_i,T_j)^{noisy}}\right| + \right. 
    \notag\\ &\hspace{-0cm} \left. W_{cor} \left|\frac{{\rm Cor}(T_i,T_j;\,\theta_i) - {\rm Cor}(T_i,T_j)^{noisy}}{{\rm Cor}(T_i,T_j)^{noisy}}\right|\right],
    \end{aligned}
\end{equation}
with, $W_{cov}$ and $W_{cor}$ verify $W_{cov} + W_{cor}=1$. The optimization is performed using SciPy's \citep{2020SciPy-NMeth} \textit{optimize} package under constraints\footnote{Many sets of tenors have been tested for the calibration, and it turned out that selecting a set of medium tenors from 3Y to 12Y led to better results for the CC. Additionally, using $W_{\text{cov}} = 3/4$ and $W_{\text{cor}} = 1/4$ gave better performance. For the initialization of the optimization process, we uniformly draw our initial parameters from the range defined in Table \ref{table:rangeparams}.}. Except for internal reports, we have not been able to find any paper in the literature dealing with this calibration procedure.

Meanwhile, for the DC, we use our already trained direct DC model, which is applied to the fifty sets of parameters. To simulate a real-life situation, the NN outputs are capped or floored when needed (e.g., if the model outputs a correlation $\rho > 1$, it will be capped at 1; for other parameters, we set floors at $10^{-8}$ if negative values are outputted). The DL model used (a classic CNN) is not expected to learn the constraints of the financial model. The caps and floors used are the same as those in the constrained optimization process of the CC.

\subsection{Results' comparison}
\label{sec:dcvscc}
\paragraph{\textbf{Computational performances}}\mbox{}\\
Once trained on our wide range of parameters (see Table \ref{table:rangeparams}), the time needed for calibration is remarkably reduced by the DC. The fifty CCs were completed in 480.1 seconds, while the DC ran in less than a second—specifically, in 0.063 seconds. This drastic reduction in processing time easily justifies the time spent beforehand to train the NN. In fact, the strong time efficiency of the DC lies in the fact that once the neural network is trained, it operates as a deterministic function, directly yielding parameters from the OQOI.

\vspace{0.5cm}
\paragraph{\textbf{Calibration accuracy}}\mbox{} \\ 
\begin{comment}
For each method, the errors are measured by the RMSE.
Overall (i-e by taking the mean of all MSEs), the DC results in a global error of 0.10 which is 2.5 lower than the 0.25 scored by the CC. \\
\\ When looking at the error for each parameter, we find the CC to be more accurate on parameters $K_x$ and $K_y$ and DC on the remaining parameters $\sigma_x$, $\sigma_y$ and $\rho$. Table \ref{table:DCvsCC} and figure \ref{fig:DCvsCC} give the MSEs obtained for each parameter for both calibrations. 
\end{comment}
We measure the error in three different ways. 
\begin{itemize}
    \item \textbf{The global error on each parameter:} We use a normalized root mean squared error (NRMSE) for each parameter. Hence, for each parameter $i \; \in \llbracket 1,\,np \rrbracket$, where $np=5$, we write:
    \begin{equation}
        \begin{aligned}
            \operatorname{NRMSE^i}&=\frac{\mathrm{RMSE^i}}{\max(\theta^i) - \min(\theta^i)}, \text{ with }   \operatorname{RMSE^i}&=\sqrt{\frac{1}{N_{sets}}\sum_{j=1}^{N_{sets}}{ (\theta^i_j - \hat{\theta^i_j})^2}}.
        \end{aligned}
    \end{equation}.

    \item \textbf{The overall error:} This measures the performance of calibrating the five parameters across the fifty sets (see step 1 of Subsection \ref{sec:compmethodo}):
    \begin{equation}
        \operatorname{Overall\;Error}= \sqrt{\sum_{i=1}^{np}{\left(\rm{NRMSE^i}\right)^2}}.
    \end{equation}
    
    \item \textbf{The local error:} This measures the global error for each of the $N_{sets} = 50$ sets of parameters. For set $j \in \, [|1,\,N_{sets}|]$, we then write:
    \begin{equation} 
        \operatorname{Local\;Error}_j=\sqrt{\sum_{i=1}^{np}\left({\frac{\theta^i_j - \hat{\theta^i_j}}{\max(\theta^i) - \min(\theta^i)}}\right)^2}.
    \end{equation}    
\end{itemize}

The global errors on each parameter and the overall error are presented in Table \ref{table:DCvsCC} and Figure \ref{fig:DCvsCC} below:
\begin{table}[H]
    \centering % used for centering table
    \begin{tabular}{c | c c c c c || c} % centered columns (5 columns)
        \hline \hline  % inserts double horizontal line
        Errors & $K_x$ & $K_y$ & $\sigma_x$ & $\sigma_y$  & $\rho$ & Overall Error \\ [1.ex] % inserts table
        \hline \hline %inserts double horizontal lines
         & & & & &\\
        CC	& 1.536	& 0.862	& 0.291	& 0.254	& 0.138 & 1.809\\ [1.ex] % inserting body of the table
        DC	& 0.0088	& 0.337	& 0.256	& 0.113	& 0.110 & 0.460\\ [1.ex] % inserting body of the
     
        \hline \hline 
    \end{tabular}
    \vspace{1.5mm}
    \caption{Global error by parameter for CC and DC, and overall errors} % title of Table
    \label{table:DCvsCC} % is used to refer this table in the text
\end{table}

We can see that the DC outperforms or is equivalent to the CC for each of the five parameters, even though the comparison parameters are outside the training sets.

\begin{figure}[H]
    \centering
    \includegraphics[width=13.5cm]{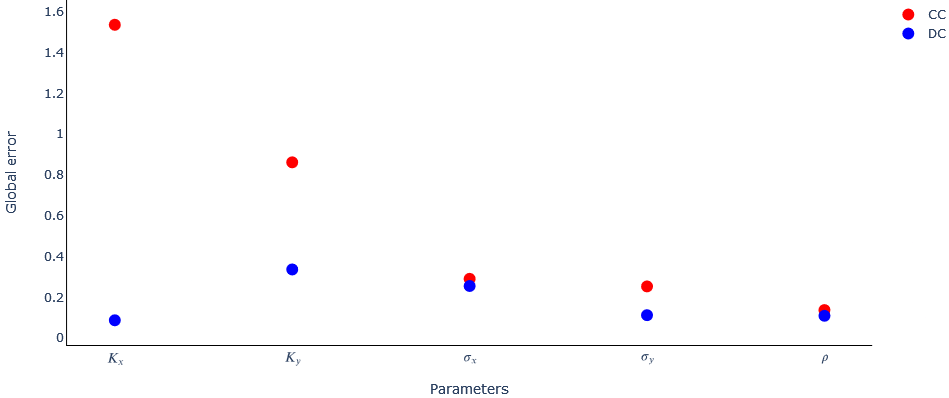}
    \vspace{-0.5cm}
    \caption{Global error by parameter for CC and DC}
    \label{fig:DCvsCC}
\end{figure}

\begin{figure}[H]
    \centering
    \includegraphics[width=13.5cm]{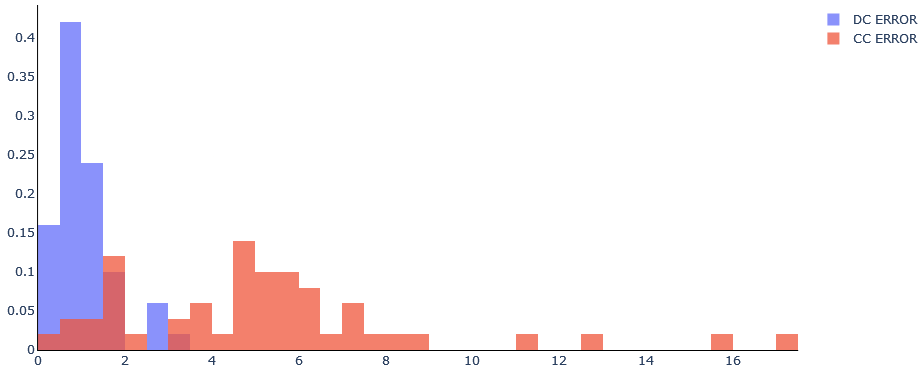}
    \vspace{-0.5cm}
    \caption{Histogram of local errors for each of the $N_{sets}$ sets}
    \label{fig:distribDCvsCC}
\end{figure}

Finally, Figure \ref{fig:distribDCvsCC} shows the distribution of local errors for each of the fifty sets. It highlights how the DC outperforms the CC in terms of global error. The DC method does not produce as many high errors as the CC.
%\vfill \mbox{}\\

\begin{comment}
We recall that as table \ref{table:res} and more generally Section \ref{sec:resec} showed, indirect DC led to, similar to better results than direct DC. We can thus infer that the indirect DC is, as well as the direct DC - as this section has shown - a valid calibration method with respect to CC.\\
\\ Similarly, choosing the right OQOI for one's given needs and preferences and applying our direct or indirect DC process (especially, the training set construction) is expected to lead to similar results.\\
\end{comment}

The previous comparisons between DC and CC were performed on a noisy synthetic data set, representative of a stressed market context. We also provide the same comparisons in a non-stressed environment, using expectations of ZC rates defined in equation \eqref{eq:expecZCs} and the related covariances and correlations. As expected, both models achieve lower errors since no noise is added. In this context, the DC still outperforms the CC. The results are presented in Table \ref{table:DCvsCC2} below. It is also interesting to note that the DC's overall error does not change much from the noisy sets, indicating that the DC calibration is more resilient to noise.

\begin{table}[h!]
    \centering % used for centering table
    \begin{tabular}{c | c c c c c || c} % centered columns (5 columns)
        \hline \hline  % inserts double horizontal line
        Errors & $K_x$ & $K_y$ & $\sigma_x$ & $\sigma_y$  & $\rho$ & Overall Error \\ [1.ex] % inserts table
        \hline \hline %inserts double horizontal lines
         & & & & &\\
        CC	& 1.288	& 0.765	& 0.290	& 0.257	& 0.135 & 1.553\\ [1.ex] % inserting body of the table
        DC	& 0.099	& 0.330	& 0.243	& 0.104	& 0.106 & 0.447\\ [1.ex] % inserting body of the
     
        \hline \hline 
    \end{tabular}
    \vspace{1.5mm}
    \caption{Global error by parameter for CC and DC, and overall errors in the noise-free environment}
    \label{table:DCvsCC2} % is used to refer this table in the text
\end{table}

\section{Beyond interest rate models calibration: CIR intensity calibration for credit risk problems}
\label{sec:apdx_CIR}

Motivated by the fact that our DC approach is designed to be systematic, meaning that it can be applied to any model as long as the conditions outlined in Section \ref{sec:Model} are met, we now apply it to a different model (the CIR intensity) and in a different context (credit risk problems). More precisely, the default intensity (DI), which can be stripped from CDS prices in a standard way (see, e.g., Chapter 22 of \cite{brigo2006interest}), is now used for another application of our indirect DC process.

\subsection{\textbf{DIs as OQOI}}\mbox{}\\
Let us consider the CIR intensity model described by the following equation:
\begin{equation}
    \label{eq:cirab}
    \begin{aligned}
            d\lambda(t) =& \left(a-b\lambda(t)\right) dt + \sigma\sqrt{\lambda(t)} \, dW_t, 
    \end{aligned}
\end{equation}
with $2a > \sigma^2$, $b > 0$, and $\lambda(0) = \lambda_0 > 0$ to ensure strictly positive DIs. Here, $(W_t)_{t \ge 0}$ denotes a Brownian motion, and $\lambda(t)$ denotes the DI at time $t$ (for more details, see, e.g., \cite{alfonsi2015affine}). In what follows, we use the indirect DC calibration procedure introduced in Section \ref{sec:DirectDC}. Only the OQOI changes from COV / COR-ZC to DIs. For this new application of our work, we calibrate only the drift parameters ($a$ and $b$). It is worth noting that maximum likelihood estimators can be built for these drift parameters (see, e.g., \cite{MR2995525, ben2013asymptotic}).

\subsection{\textbf{Data set construction}}\mbox{}\\
Once again, the process remains the same as the one described in Subsection \ref{sec:datasetconstr} as we still use synthetic data sets, except for the third step, where we apply the drift implicit discretization scheme to the stochastic differential equation. More precisely, we consider a uniform grid $t_i = i T/n, \; n \in \mathbb{N}^*$, and for $4a > \sigma^2$ and $T/n < 2b$, the drift implicit scheme introduced by \citep{alfonsi2005discretization} is given by:
\begin{equation}
    \label{eq:disc}
    \lambda_{t_{i+1}}=\left(\frac{\frac{\sigma}{2}\left(W_{t_{i+1}}-W_{t_{i}}\right)+\sqrt{\lambda_{t_{i}}}+\sqrt{\left(\frac{\sigma}{2}\left(W_{t_{i+1}}-W_{t_{i}}\right)+\sqrt{\lambda_{t_{i}}}\right)^{2}+4\left(1+\frac{b T}{2 n}\right) \frac{a-\sigma^{2} / 4}{2} \frac{T}{n}}}{2\left(1+\frac{b T}{2 n}\right)}\right)^{2}.
\end{equation}
Note that, we could have also used the actual distribution of a solution of the CIR model (see for instance, \cite{CIR_base}).

We define for each of the parameters a range of 3 equidistant values: from 0 to 5 for $b$, 0\% to 5\% for $a$, and 0.1\% to 3\% for $\sigma$. We then take all combinations of values verifying $2a > \sigma^2$, which, since $a > 0$, ensures $4a > \sigma^2$. Using equation \eqref{eq:disc}, we compute a thousand paths of DIs for each set of parameters. We keep the same time range as previously: 105 steps of size 1 week, which is about 2 years. The value of $\lambda_0$ is set to 0.5\%. Similarly to ZC rates, DI simulations lead to many random paths. We can either consider one typical path that would be representative of all the others (as we do in Subsection \ref{sec:zcasoqoi} by taking the expectation of ZCs) or consider the entire set of simulated paths. In what follows, we use the latter option to illustrate this alternative approach, as in Subsection \ref{sec:zcasoqoi} we applied the first solution. This approach leads to a dataset of 18,000 entries with 107 features (106 time steps and $\sigma$). Applying an 80\%-20\% split results in training and test sets of 14,400 and 3,600 entries, respectively.

\subsection{The Neural Network Architecture}\mbox{}\\
%\vspace{-1.cm}
The architecture is similar to the indirect DC's; we have a shallow FCN with only 3 layers in this case. The input layer has 106 neurons ($nf$, the number of features), the hidden layer has 1,500 neurons ($h$), and the output layer has 2 neurons ($np$, for the 2 parameters $a$ and $b$ to calibrate). The activation functions are ReLUs, except for the prediction layer, which does not have an activation function. The Adam optimizer is used to minimize the MSE loss with a learning rate of 0.001. The training is performed over 1,000 epochs with mini-batches of size 256. The architecture can be represented as follows:
\vspace{-0.1cm}
% NEURAL NETWORK
%h!
\begin{figure}[H]
    \centering
    \begin{tikzpicture}[x=3.3cm, y=0.85cm]
      \readlist\Nnod{3,4,2} % array of number of nodes per layer
      \readlist\Nstr{nf, h, np} % array of string number of nodes per layer
      \readlist\Cstr{u,h,v} % array of coefficient symbol per layer
      \def\yshift{0.55} % shift last node for dots
      
      % LOOP over LAYERS
      \foreachitem \N \in \Nnod{
        \def\lay{\Ncnt} % alias of index of current layer
        \pgfmathsetmacro\prev{int(\Ncnt-1)} % number of previous layer
        \foreach \i [evaluate={\c=int(\i==\N); 
                               \y=\N/2-\i-\c*\yshift;
                               \x=\lay; 
                               \n=\nstyle;
                               \index=(\i<\N?int(\i):"\Nstr[\Ncnt]");}] in {1,...,\N}{ % loop over nodes
                               %\index="\Nstr[\N]";}] in {1,...,\N}{ % loop over nodes
          % NODES
          \node[node \n] (N\lay-\i) at (\x,\y) {$\strut\Cstr[\n]_{\index}$};
          
          % CONNECTIONS
          \ifnumcomp{\lay}{>}{1}{ % connect to previous layer
            \foreach \j in {1,...,\Nnod[\prev]}{ % loop over nodes in previous layer
              \draw[-stealth] (N\prev-\j) -- (N\lay-\i);
              \draw[-stealth] (N\prev-\j) -- (N\lay-\i);
            }
            \ifnum \lay=\Nnodlen
              \draw[-stealth] (N\lay-\i) --++ (0.5,0); % arrows out
            \fi
          }
          {
            \draw[-stealth] (0.5,\y) -- (N\lay-\i); % arrows in
          }
        }
        
        \path (N\lay-\N) --++ (0,1+\yshift) node[midway,scale=1.6] {$\vdots$}; % dots
      }
      
      % LABELS
      \node[above=3,align=center,mydarkgreen] at (N1-1.90) {Input\\[-0.2em]layer};
      \node[above=2,align=center,mydarkblue] at (N2-1.90) {Hidden\\[-0.2em]layers};
      \node[above=3,align=center,mydarkred] at (N\Nnodlen-1.90) {Output/Prediction \\[-0.2em]layer};
    \end{tikzpicture}

    \vspace{-0.2cm}
    \caption{FCN Architecture of the indirect DC for the CIR intensity model}
\end{figure}

\subsection{Calibration Results}
\begin{figure}[H]
    \centering
    \includegraphics[width=15.cm, height=6.5cm]{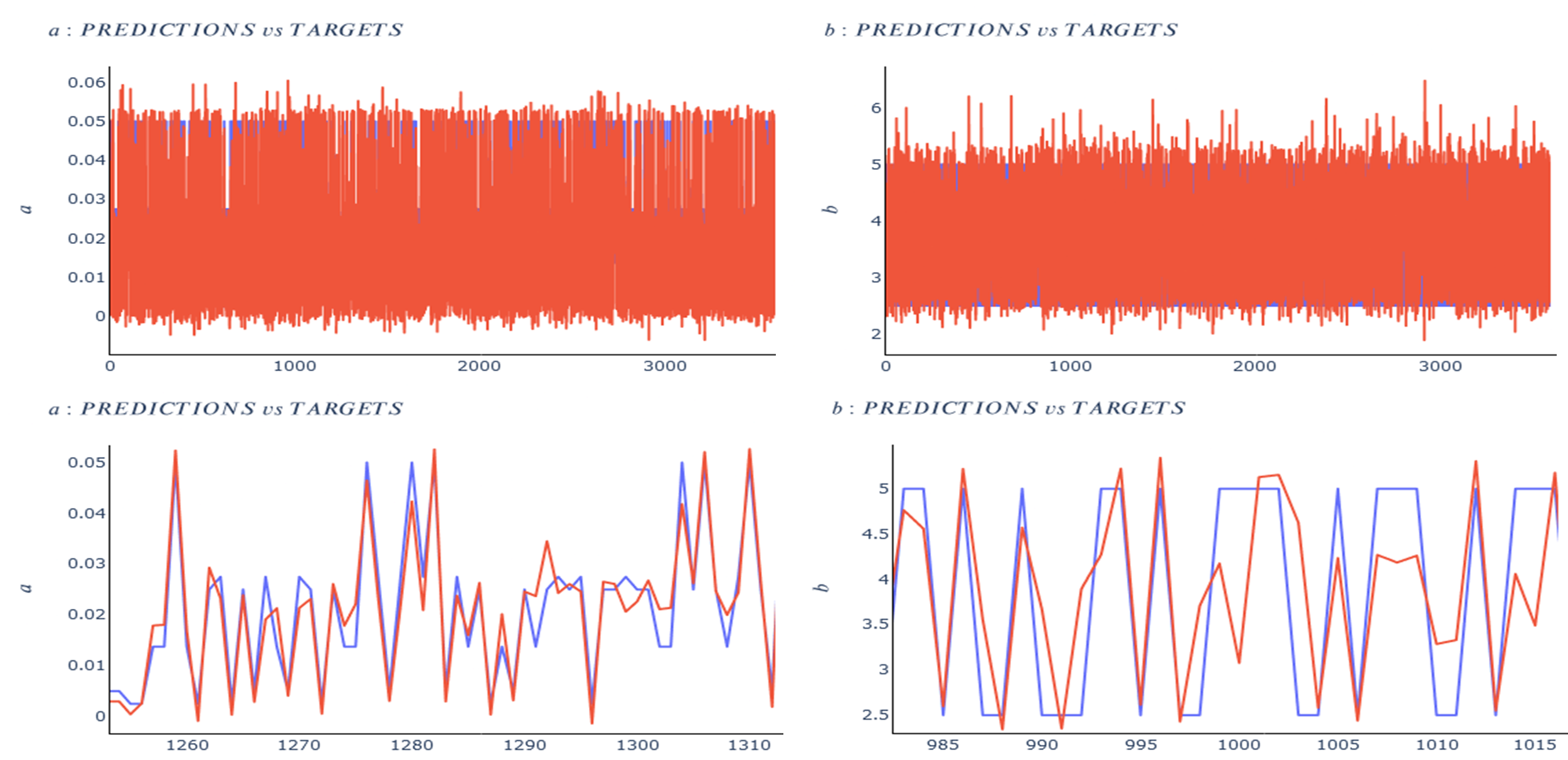} % second figure itself
    \vspace{-0.25cm}
    \caption{Indirect DC on the CIR intensity model: Fitting Curves with and without zoom on parameters $a$ and $b$. The red curves are predictions and the blue curves actual values.}
    \label{fig:fitcir}
\end{figure}

\mbox{} \\
Both accuracy and time efficiency are good, as calibration on the test set is performed in under a second. The MSE on $a$ is $2 \times 10^{-5}$ and $0.5776$ on $b$. Figure \ref{fig:fitcir} shows the fitting curves.
%\vspace{-.75cm}
% \begin{figure}[H]
%     \centering
%     \includegraphics[width=0.7\textwidth, height=3.3cm]{_V1_a.png}%\hfill
%     \vspace{-0.7cm}
%     \includegraphics[width=0.7\textwidth, height=3.3cm]{_V1_a_zoom.png}%\hfill
%     \vspace{-0.7cm}
%     \includegraphics[width=0.7\textwidth, height=3.3cm]{_V1_b.png}%\hfill
%     \vspace{-0.7cm}
%     \includegraphics[width=0.7\textwidth, height=3.3cm]{_V1_b_zoom.png}%\hfill
%     \caption{Indirect DC on the CIR intensity model: Fitting Curves with and without zoom on parameters $a$ and $b$}
%     \label{fig:fitcir}
% \end{figure}

\section{Conclusion}

Despite the growing use of deep learning and related artificial intelligence techniques in finance, analytical models, such as interest rate models like the G2++, remain commonly used due to the expertise that professionals and researchers have developed over the last few decades. However, this does not mean that DL and AI, in general, should not be leveraged to improve our use of these models. In this paper, we have demonstrated that DL can be effectively used to calibrate IR and intensity models. Starting with parameters from a classic market calibration of the G2++ model, we generated a synthetic dataset of parameters by uniformly sampling around those reference parameters. Using results from the literature and our reference parameters, we generated two types of datasets. The first, for the so-called indirect DC, consists of correlations and covariances of ZC and FWD rates. The second set serves the direct DC and consists of ZC and FWD rate curves. For the first model, we used a shallow fully connected network and for the second, a shallow convolutional neural network. We have shown that covariances provided more information to the neural networks, and, as a consequence, the indirect DC using correlations made about double the errors of the covariance-based indirect DC. The direct DC with ZC rates, like the indirect DC with covariances, achieved good accuracy. The errors and computational performance obtained allow us to conclude that using DL leads to very fast calibration with low errors for the G2++ model. Our DC approach is designed to be systematic, meaning that it can be applied to any model with data fitting our requirements for the observable quantity of interest, i.e., it must be observable in the market and have an analytical expression. Indeed, we achieved a global error significantly lower than the classical optimization method in less than a second, compared to the 480 seconds required by the classic method. Additionally, as we intended, the models are straightforward to use and require only easily available data with minimal to no transformation. As a next step, we could improve the results of the indirect DC by focusing on the covariances, allowing more tenors, and applying dimension reduction techniques, such as auto-encoders or Principal Component Analysis (PCA). Additionally, deepening the network could also enhance performance, but this might require more complex architectures to avoid overfitting, such as those incorporating regularization techniques. It would be of high interest to apply our pretrained NNs to other models, such as option pricing models or those with path-dependent parameters. This could be the focus of future work.

\newpage 

\bibliographystyle{apalike}
\bibliography{references}  

\begin{thebibliography}{}

\bibitem[Albawi et~al., 2017]{albawi2017understanding}
Albawi, S., Mohammed, T.~A., and Al-Zawi, S. (2017).
\newblock Understanding of a convolutional neural network.
\newblock In {\em 2017 International Conference on Engineering and Technology
  (ICET)}, pages 1--6. Ieee.

\bibitem[Alfonsi, 2005]{alfonsi2005discretization}
Alfonsi, A. (2005).
\newblock On the discretization schemes for the cir (and bessel squared)
  processes.

\bibitem[Alfonsi et~al., 2015]{alfonsi2015affine}
Alfonsi, A. et~al. (2015).
\newblock {\em Affine diffusions and related processes: simulation, theory and
  applications}, volume~6.
\newblock Springer.

\bibitem[Ben~Alaya and Kebaier, 2012]{MR2995525}
Ben~Alaya, M. and Kebaier, A. (2012).
\newblock Parameter estimation for the square-root diffusions: ergodic and
  nonergodic cases.
\newblock {\em Stoch. Models}, 28(4):609--634.

\bibitem[Ben~Alaya and Kebaier, 2013]{ben2013asymptotic}
Ben~Alaya, M. and Kebaier, A. (2013).
\newblock Asymptotic behavior of the maximum likelihood estimator for ergodic
  and nonergodic square-root diffusions.
\newblock {\em Stochastic Analysis and Applications}, 31(4):552--573.

\bibitem[Bloch, 2019]{bloch2019neural}
Bloch, D.~A. (2019).
\newblock Neural networks based dynamic implied volatility surface.
\newblock {\em Available at SSRN 3492662}.

\bibitem[Brigo and Mercurio, 2006]{brigo2006interest}
Brigo, D. and Mercurio, F. (2006).
\newblock {\em Interest rate models-theory and practice: with smile, inflation
  and credit}, volume~2.
\newblock Springer.

\bibitem[B{\"u}chel et~al., 2022]{buchel2022deep}
B{\"u}chel, P., Kratochwil, M., Nagl, M., and R{\"o}sch, D. (2022).
\newblock Deep calibration of financial models: turning theory into practice.
\newblock {\em Review of Derivatives Research}, pages 1--28.

\bibitem[Buehler et~al., 2019]{buehler2019deep}
Buehler, H., Gonon, L., Teichmann, J., and Wood, B. (2019).
\newblock Deep hedging.
\newblock {\em Quantitative Finance}, 19(8):1271--1291.

\bibitem[Chen et~al., 2020]{chen2020deep}
Chen, L., Pelger, M., and Zhu, J. (2020).
\newblock Deep learning in asset pricing.
\newblock {\em Available at SSRN 3350138}.

\bibitem[Covitz and Downing, 2007]{covitz2007liquidity}
Covitz, D. and Downing, C. (2007).
\newblock Liquidity or credit risk? the determinants of very short-term
  corporate yield spreads.
\newblock {\em The Journal of Finance}, 62(5):2303--2328.

\bibitem[Cox et~al., 1985]{CIR_base}
Cox, J.~C., Ingersoll, J.~E., and Ross, S.~A. (1985).
\newblock A theory of the term structure of interest rates.
\newblock {\em Econometrica}, 53(2):385--407.

\bibitem[Creswell et~al., 2018]{creswell2018generative}
Creswell, A., White, T., Dumoulin, V., Arulkumaran, K., Sengupta, B., and
  Bharath, A.~A. (2018).
\newblock Generative adversarial networks: An overview.
\newblock {\em IEEE Signal Processing Magazine}, 35(1):53--65.

\bibitem[Gurrieri et~al., 2009]{gurrieri2009calibration}
Gurrieri, S., Nakabayashi, M., and Wong, T. (2009).
\newblock Calibration methods of hull-white model.
\newblock {\em Available at SSRN 1514192}.

\bibitem[Heaton et~al., 2016]{heaton2016deep}
Heaton, J., Polson, N.~G., and Witte, J.~H. (2016).
\newblock Deep learning in finance.
\newblock {\em arXiv preprint arXiv:1602.06561}.

\bibitem[Heaton et~al., 2017]{heaton2017deep}
Heaton, J.~B., Polson, N.~G., and Witte, J.~H. (2017).
\newblock Deep learning for finance: deep portfolios.
\newblock {\em Applied Stochastic Models in Business and Industry},
  33(1):3--12.

\bibitem[Hernandez, 2016]{hernandez2016model}
Hernandez, A. (2016).
\newblock Model calibration with neural networks.
\newblock {\em Available at SSRN 2812140}.

\bibitem[Hochreiter and Schmidhuber, 1997]{hochreiter1997long}
Hochreiter, S. and Schmidhuber, J. (1997).
\newblock Long short-term memory.
\newblock {\em Neural computation}, 9(8):1735--1780.

\bibitem[Horvath et~al., 2021]{horvath2021deep}
Horvath, B., Muguruza, A., and Tomas, M. (2021).
\newblock Deep learning volatility: a deep neural network perspective on
  pricing and calibration in (rough) volatility models.
\newblock {\em Quantitative Finance}, 21(1):11--27.

\bibitem[Huang et~al., 2020]{huang2020deep}
Huang, J., Chai, J., and Cho, S. (2020).
\newblock Deep learning in finance and banking: A literature review and
  classification.
\newblock {\em Frontiers of Business Research in China}, 14:1--24.

\bibitem[Hull and White, 2001]{hull2001general}
Hull, J. and White, A. (2001).
\newblock The general hull--white model and supercalibration.
\newblock {\em Financial Analysts Journal}, 57(6):34--43.

\bibitem[Jacovides, 2008]{jacovides2008forecasting}
Jacovides, A. (2008).
\newblock {\em Forecasting Interest Rates from the Term Structure: Support
  Vector Machines Vs Neural Networks}.
\newblock PhD thesis, Citeseer.

\bibitem[Jang et~al., 2021]{jang2021deepoption}
Jang, J.~H., Yoon, J., Kim, J., Gu, J., and Kim, H.~Y. (2021).
\newblock Deepoption: A novel option pricing framework based on deep learning
  with fused distilled data from multiple parametric methods.
\newblock {\em Information Fusion}, 70:43--59.

\bibitem[Kaelbling et~al., 1996]{kaelbling1996reinforcement}
Kaelbling, L.~P., Littman, M.~L., and Moore, A.~W. (1996).
\newblock Reinforcement learning: A survey.
\newblock {\em Journal of artificial intelligence research}, 4:237--285.

\bibitem[Kingma and Ba, 2014]{kingma2014adam}
Kingma, D.~P. and Ba, J. (2014).
\newblock Adam: A method for stochastic optimization.
\newblock {\em arXiv preprint arXiv:1412.6980}.

\bibitem[Klad{\'\i}vko, 2007]{kladivko2007maximum}
Klad{\'\i}vko, K. (2007).
\newblock Maximum likelihood estimation of the cox-ingersoll-ross process: the
  matlab implementation.
\newblock {\em Technical Computing Prague}, 7(8).

\bibitem[Liu et~al., 2019]{liu2019neural}
Liu, S., Borovykh, A., Grzelak, L.~A., and Oosterlee, C.~W. (2019).
\newblock A neural network-based framework for financial model calibration.
\newblock {\em Journal of Mathematics in Industry}, 9:1--28.

\bibitem[Mbaye and Vrins, 2022]{mbaye2022affine}
Mbaye, C. and Vrins, F. (2022).
\newblock Affine term structure models: A time-change approach with perfect fit
  to market curves.
\newblock {\em Mathematical Finance}, 32(2):678--724.

\bibitem[Ng et~al., 2011]{ng2011sparse}
Ng, A. et~al. (2011).
\newblock Sparse autoencoder.
\newblock {\em CS294A Lecture notes}, 72(2011):1--19.

\bibitem[Oh and Han, 2000]{oh2000using}
Oh, K.~J. and Han, I. (2000).
\newblock Using change-point detection to support artificial neural networks
  for interest rates forecasting.
\newblock {\em Expert systems with applications}, 19(2):105--115.

\bibitem[Orlando et~al., 2019]{orlando2019interest}
Orlando, G., Mininni, R.~M., and Bufalo, M. (2019).
\newblock Interest rates calibration with a cir model.
\newblock {\em The Journal of Risk Finance}.

\bibitem[Pal and Mitra, 1992]{pal1992multilayer}
Pal, S.~K. and Mitra, S. (1992).
\newblock Multilayer perceptron, fuzzy sets, classifiaction.

\bibitem[Pironneau, 2019]{pironneau2019calibration}
Pironneau, O. (2019).
\newblock Calibration of heston model with keras.

\bibitem[Russo and Fabozzi, 2017]{russo2017calibrating}
Russo, V. and Fabozzi, F.~J. (2017).
\newblock Calibrating short interest rate models in negative rate environments.
\newblock {\em The Journal of Derivatives}, 24(4):80--92.

\bibitem[Russo and Torri, 2019]{russo2019calibration}
Russo, V. and Torri, G. (2019).
\newblock Calibration of one-factor and two-factor hull--white models using
  swaptions.
\newblock {\em Computational Management Science}, 16(1):275--295.

\bibitem[Schlenkrich, 2012]{schlenkrich2012efficient}
Schlenkrich, S. (2012).
\newblock Efficient calibration of the hull white model.
\newblock {\em Optimal Control Applications and Methods}, 33(3):352--362.

\bibitem[Vasicek, 1977]{vasicek1977equilibrium}
Vasicek, O. (1977).
\newblock An equilibrium characterization of the term structure.
\newblock {\em Journal of financial economics}, 5(2):177--188.

\bibitem[Vela, 2013]{vela2013forecasting}
Vela, D. (2013).
\newblock Forecasting latin-american yield curves: An artificial neural network
  approach.
\newblock {\em Borradores de Econom{\'\i}a; No. 761}.

\bibitem[Virtanen et~al., 2020]{2020SciPy-NMeth}
Virtanen, P., Gommers, R., Oliphant, T.~E., Haberland, M., Reddy, T.,
  Cournapeau, D., Burovski, E., Peterson, P., Weckesser, W., Bright, J., {van
  der Walt}, S.~J., Brett, M., Wilson, J., Millman, K.~J., Mayorov, N., Nelson,
  A. R.~J., Jones, E., Kern, R., Larson, E., Carey, C.~J., Polat, {\.I}., Feng,
  Y., Moore, E.~W., {VanderPlas}, J., Laxalde, D., Perktold, J., Cimrman, R.,
  Henriksen, I., Quintero, E.~A., Harris, C.~R., Archibald, A.~M., Ribeiro,
  A.~H., Pedregosa, F., {van Mulbregt}, P., and {SciPy 1.0 Contributors}
  (2020).
\newblock {{SciPy} 1.0: Fundamental Algorithms for Scientific Computing in
  Python}.
\newblock {\em Nature Methods}, 17:261--272.

\end{thebibliography}

\newpage

\appendix

%%%%%%%%%%%%%%%%%%%%%%%%%%%%%%%%%
%Fii la yamm. Mais il faut changer figures 1 et 2 et 3 mettre légende en bas passer au fond blanc 
% Must change graph civ coir (xtick)
% GRAPH ZC UGLY AF
%%%%%%%%%%%%%%%%%%%%%%%%%%%%%%%%%
\section{Market ZC Correlations and Correlations}
\label{sec:apdx_mrkt_covcor}

We can observe ZC rates' covariance and correlation curves similar to those simulated for our synthetic data sets\footnote{The data used to obtain the CORs and COVs can be found at \url{https://data.nasdaq.com/data/FED/SVENY-us-treasury-zerocoupon-yield-curve}.}.

\vspace{-5mm}
\begin{figure}[H]
    \centering
    \begin{minipage}{0.475\textwidth}
        \begin{figure}[H]
            \centering
            \includegraphics[width=7.5cm, height=6cm]{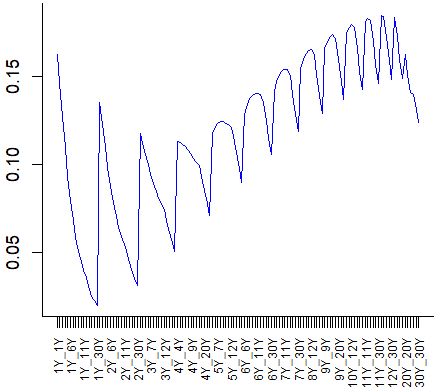} % first figure itself
            \vspace{-8.5mm}
            \caption{\footnotesize Covariance curve of the US Treasury ZC rate curve; the data used ranges from January 2020 to October 2021.}
            \label{fig:fig17}
        \end{figure}
    \end{minipage}\hfill
    \begin{minipage}{0.475\textwidth}
        \begin{figure}[H]
            \centering
            \includegraphics[width=7.5cm, height=6cm]{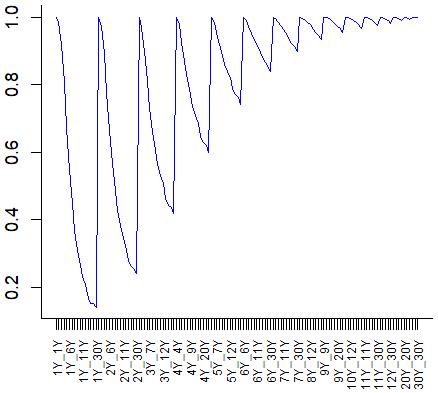} % second figure itself
            \vspace{-8.5mm}
            \caption{\footnotesize Correlation curve of the US Treasury ZC rate curve; the data used ranges from January 2020 to October 2021.}
            \label{fig:fig171}
        \end{figure}
    \end{minipage}
\end{figure}

\section{Market ZC Curve}
\label{sec:apdx_ZC_gen}
We also observe ZC rates' graphs similar to those simulated for our synthetic data set\footnote{The data can be retrieved from \url{https://www.ecb.europa.eu/stats/financial_markets_and_interest_rates/euro_area_yield_curves/html/index.it.html}.}.

\begin{figure}[H]
    \centering
    \includegraphics[height=7.cm, width = 11cm]{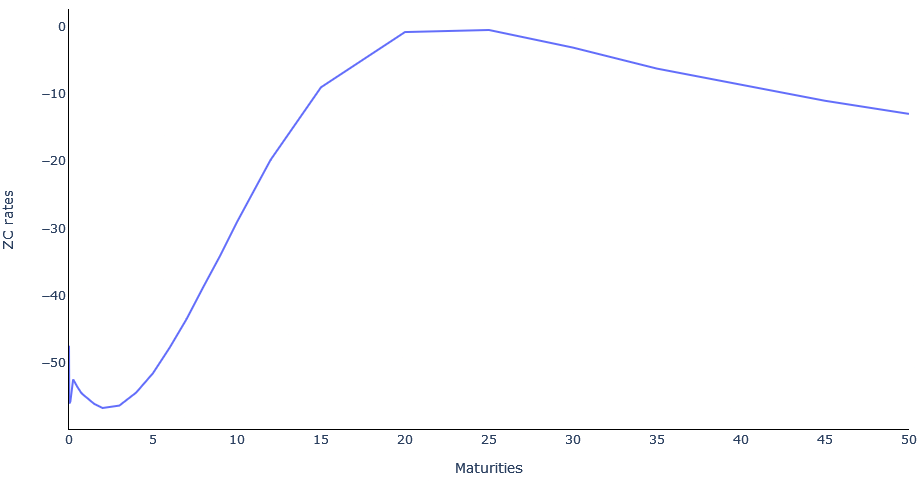} % first figure itself
    \caption{Initial Euro ZC rates in basis points as of 2020-11-04}
    \label{fig:mkt0}
\end{figure}

\begin{comment}
\begin{figure}[H]
    \centering
    \includegraphics[clip, trim=1cm 2cm 1cm 1cm, width=1.00\textwidth]{ZC_curve.pdf} 
    \caption{ZC Curve of the US treasury}
    \label{fig:fig18}
\end{figure}
\end{comment}

\section{Unfeasible backpropagation theorem's proof}
\label{sec:proof}
\begin{comment}
\paragraph{Theorem: } (Unfeasible backpropagation)
\emph{In a feed forward neural network architecture, if the derivatives of the inputs w.r.t the targets equal 0, then so does the derivatives of the loss function w.r.t to the weights, thus preventing the gradient descent by backpropagation and subconsequently stalling the training.}
\end{comment}

%--------------------------------------------------------------------------

%--------------------------------------------------------------------------
\begin{proof}[Proof of Theorem \ref{theo:unfeasbackprop}.]
Without loss of generality, we consider an unbiased NN with inputs and outputs of dimension 1 and with one hidden layer. The NN is described in Figure \ref{fig:nnprf}, where \( u \in \mathbb{R} \) is the input, \( h \) is the result of applying the hidden layer's activation function to \( u \), and \( y \) is the output of the NN.

% NEURAL NETWORK
\begin{figure}[h!]
    \centering
    \begin{tikzpicture}[x=3.3cm,y=1.6cm]
      \readlist\Nnod{1,1,1} % array of number of nodes per layer
      \readlist\Nstr{, , } % array of string number of nodes per layer
      \readlist\Cstr{u,h,y} % array of coefficient symbol per layer
      \def\yshift{0.55} % shift last node for dots
      
      % LOOP over LAYERS
      \foreachitem \N \in \Nnod{
        \def\lay{\Ncnt} % alias of index of current layer
        \pgfmathsetmacro\prev{int(\Ncnt-1)} % number of previous layer
        \foreach \i [evaluate={\c=int(\i==\N); 
                               \y=\N/2-\i-\c*\yshift;
                               \x=\lay; 
                               \n=\nstyle;
                               \index=(\i<\N?int(\i):"\Nstr[\Ncnt]");}] in {1,...,\N}{ % loop over nodes
                               %\index="\Nstr[\N]";}] in {1,...,\N}{ % loop over nodes
          % NODES
          \node[node \n] (N\lay-\i) at (\x,\y) {$\strut\Cstr[\n]_{\index}$};
          
          % CONNECTIONS
          \ifnumcomp{\lay}{>}{1}{ % connect to previous layer
            \foreach \j in {1,...,\Nnod[\prev]}{ % loop over nodes in previous layer
              \draw[-stealth] (N\prev-\j) -- (N\lay-\i);
              \draw[-stealth] (N\prev-\j) -- (N\lay-\i);
            }
            \ifnum \lay=\Nnodlen
              \draw[-stealth] (N\lay-\i) --++ (0.5,0); % arrows out
            \fi
          }
          {
            \draw[-stealth] (0.5,\y) -- (N\lay-\i); % arrows in
          }
        }
      }
      
      % LABELS
      \node[above=3,align=center,mydarkgreen] at (N1-1.90) {Input\\[-0.2em]layer};
      \node[above=2,align=center,mydarkblue] at (N2-1.90) {Hidden\\[-0.2em]layer};
      \node[above=3,align=center,mydarkred] at (N\Nnodlen-1.90) {Output\\[-0.2em]layer};
    \end{tikzpicture}

    \caption{Simplified feed-forward neural network }
    \label{fig:nnprf}
 
\end{figure}
Let \( w = \left(w_0, \; w_1\right) \in \mathbb{R}^2 \) define the weights, where \( w_0 \) is the weight on the hidden layer and \( w_1 \) is the weight on the output layer. These unbiased weights are updated between optimization steps \( k \) and \( k+1 \) by gradient descent: \( w^{k+1} = w^{k} - \varepsilon \nabla_{w}L(w^k) \), \( k \in \mathbb{N} \), where \( L \) is any smooth loss function and \( \varepsilon \) is the learning rate. The backpropagation uses the chain rule and is written as:
\begin{equation}
    \label{eq:prf1}
    \begin{aligned}
        \frac{\partial L}{\partial w_1}& = \frac{\partial L}{\partial y} \frac{\partial y}{\partial w_1}.
    \end{aligned}
\end{equation}
 
Let us focus on \( \frac{\partial L}{\partial y} \). 
Let \( f \) be the function linking the inputs \( u \) of the neural network to the outputs \( y \).
%We take for example the MSE, $t$ being the target, $L(f(u), \; t) = \left(f(u) - t\right)^2$. 
The important step is to use the fact that our inputs are functions of the targets (and consequently of the outputs). In fact, \( u = u(y) \) and by the theorem's hypothesis, \( \frac{\partial u}{\partial y} = 0 \). So we have:
\begin{equation}
    \label{eq:eq23}
    \frac{\partial L}{\partial y} = \frac{\partial L}{\partial f}\frac{\partial f}{\partial y} =\frac{\partial L}{\partial f}\frac{\partial f}{\partial u}\underbrace{\frac{\partial u}{\partial y}}_{=0} = 0.
\end{equation}

\begin{comment}
\begin{equation}
    \label{eq:eq22}
    \begin{aligned}
        \frac{\partial L}{\partial y} &= \frac{\partial (f(u(y)) - t)^2}{\partial y}
        & = 2(f(u(y)) - t)\frac{\partial (f(u(y)) - t)}{\partial y}
        & = 2(f(u(y)) - t)\frac{\partial f}{\partial u}\underbrace{\frac{\partial u(y)}{\partial y}}_{=0} = 0. \\
    \end{aligned}
\end{equation}
\end{comment}

Hence, by \eqref{eq:prf1} $\frac{\partial L}{\partial w_1} = 0$. On the other hand, when backpropagating to $w_0$ as we will write, $\frac{\partial L}{\partial w_0} = \frac{\partial L}{\partial y} \frac{\partial y}{\partial h}\frac{\partial h}{\partial w_0}$ and we have already shown $\frac{\partial L}{\partial y}=0$, then we also deduce $\frac{\partial L}{\partial w_0} = 0$. Consequently we have $\nabla_{w}L(w^k) = 0_{\mathbb{R}^2} \text{ which implies that } w^{k+1}\;=\;w^{k}, \; k \in \mathbb{N}$.

These arguments can easily be generalized for more complex architectures, since the basic principle of the backpropagation which is to start from the rightmost partial derivative and propagate through, still remains. Hence, for outputs $(y_i)_{i \in [|1,\;d|]}$ of dimension $d \ge 1$, the only requirement is that all the partial derivatives of the inputs w.r.t to the outputs must be null, that is $\forall i \in[|1,d|], \frac{\partial u}{\partial y_i}=0$.

These arguments can easily be generalized for more complex architectures, since the basic principle of backpropagation, which is to start from the rightmost partial derivative and propagate through, still remains. Hence, for outputs $ (y_i)_{i \in \llbracket 1,\;\operatorname{dim} \rrbracket} $ of dimension $ \operatorname{dim} \ge 1 $, the only requirement is that all the partial derivatives of the inputs with respect to the outputs must be zero, that is $$ \forall i \in \llbracket 1,\operatorname{dim} \rrbracket, \frac{\partial u}{\partial y_i} = 0. $$

\end{proof}
%--------------------------------------------------------------------------

\end{document}